\documentclass{article}
\usepackage[a4paper, margin=1.5in]{geometry}

\usepackage{pifont}
\usepackage{subcaption}
\usepackage[table]{xcolor}
\usepackage{tcolorbox}
\usepackage{hyperref}
\usepackage{enumitem}

\usepackage{acronym}
\usepackage{listing}
\usepackage{multirow}
\usepackage{booktabs}
\usepackage{amsmath}
\usepackage{amsfonts}

\usepackage{tikz}
\usetikzlibrary{positioning,arrows.meta,calc}

\newcommand{\rowgray}{\rowcolor{gray!10}}
\newcommand{\cmark}{\ding{51}}
\newcommand{\xmark}{\ding{55}}
\newcommand{\best}[1]{\cellcolor{green!15}\textbf{#1}}
\newcommand{\secondbest}[1]{\cellcolor{yellow!25}{#1}}
\usepackage{adjustbox}
\newcommand{\hl}[1]{\textcolor{black}{#1}}

\acrodef{RS}{recommender system}
\acrodef{LLM}{large language model}
\acrodef{ST}{Sentence-Transformers}
\acrodef{CF}{Collaborative Filtering}
\acrodef{PCA}{Principal Component Analysis}
\acrodef{RAG}{retrieval-augmented generation}
\acrodef{CCA}{Canonical Correlation Analysis}
\newcommand{\RAGBench}{\textsc{RAG-VisualRec}}

\title{\textsc{RAG-VisualRec:} An Open Resource for Vision- and Text-Enhanced Retrieval-Augmented Generation in Recommendation}

\author{
    Ali Tourani\footnote{Interdisciplinary Centre for Security, Reliability, and Trust (SnT), University of Luxembourg, Luxembourg. Institute for Advanced Studies, University of Luxembourg, Luxembourg. \texttt{ali.tourani@uni.lu}},
    Fatemeh Nazary\footnote{Polytechnic University of Bari, Bari, Italy. \texttt{fatemeh.nazary@poliba.it}},
    Yashar Deldjoo\footnote{Polytechnic University of Bari, Bari, Italy. \texttt{deldjooy@acm.org}}
}

\begin{document}

\maketitle

\begin{abstract}
This paper addresses the challenge of building multimodal recommender systems for the movie domain, where sparse item metadata (e.g., title and genres) can limit retrieval quality and downstream recommendations. We introduce \textsc{RAG-VisualRec}, an open resource and reproducible pipeline that combines (i) LLM-generated item-side plot descriptions and (ii) trailer-derived visual (and optional audio) embeddings, supporting both retrieval-augmented generation (RAG) and collaborative-filtering style workflows. Our pipeline augments sparse metadata into richer textual signals and integrates modalities via configurable fusion strategies (e.g., PCA and CCA) before retrieval and optional LLM-based re-ranking.
Beyond providing the resource, we provide a complementary analysis that increases transparency and reproducibility. In particular, we introduce \texttt{LLMGenQC}, a critic-based quality-control module (LLM-as-judge) that audits synthetic synopses for semantic alignment with metadata, consistency, safety, and basic sanity checks, releasing critic scores and pass/fail labels alongside the generated artifacts. We report ablation studies that quantify the impact of key design choices, including retrieval depth, fusion strategy, and user-embedding construction. Across experiments, CCA-based fusion consistently improves recall over unimodal baselines, while LLM-based re-ranking typically improves nDCG by refining top-$K$ selection from the retrieved candidate pool, especially when textual evidence is limited. By releasing \textsc{RAG-VisualRec}, we enable further research on multimodal RAG recommenders, quality auditing of LLM-generated side information, and long-tail oriented evaluation protocols. All code, data, and detailed documentation are publicly available at:
\url{https://github.com/RecSys-lab/RAG-VisualRec}.
\end{abstract}

\section{Introduction}
\label{sec:intro}

\begin{figure}[t!]
     \centering
     \includegraphics[width=1.0\textwidth]{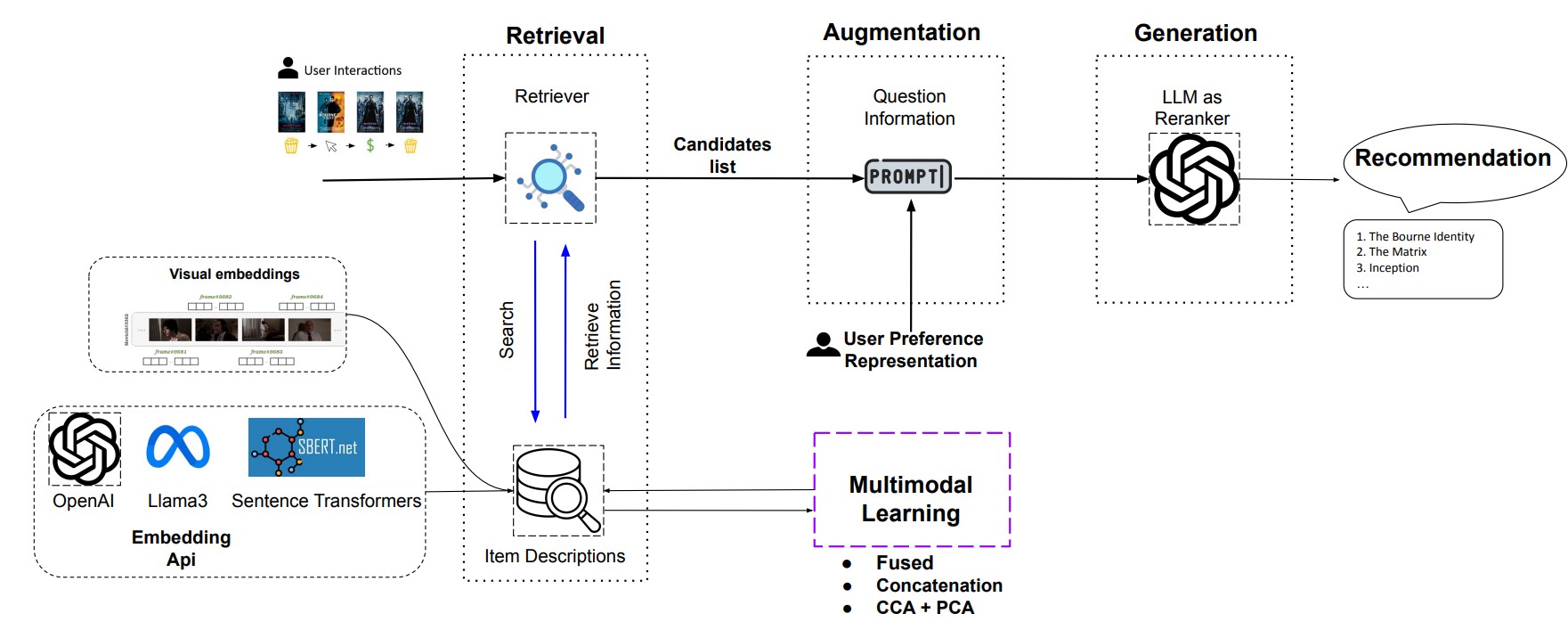}
     \caption{The overall pipeline of the proposed multimodal RAG framework.}
     \label{fig:overview}
\end{figure}

Recommender systems (RS) increasingly integrate large language models (LLMs) to interpret user intent, capture fine-grained item semantics, and generate explanations beyond interaction-only personalization.
Retrieval-augmented generation (RAG) strengthens this paradigm by grounding generation in retrieved evidence, mitigating hallucination and stale knowledge, and improving robustness when item-side metadata is sparse or noisy~\cite{gao2023retrieval,zeng2024federated,deldjoo2024recommendation,wu2023ruel}. A standard RAG pipeline (Fig.~\ref{fig:overview}) includes three stages: (1) \emph{retrieval} (select candidate items or evidence), (2) \emph{augmentation} (condition the prompt on retrieved context), and (3) \emph{generation} (produce the final recommendation list and rationale). While collaborative filtering (\ac{CF}) remains a strong retrieval component, embedding-based retrieval is often preferred in RAG settings because it unifies heterogeneous signals (text, vision, audio) in a shared vector space, enabling faster adaptation to catalogue changes and improved exposure of long-tail or newly introduced items. For instance, if a film suddenly wins an award, a RAG recommender can ingest the new fact into its vector store and refresh its suggestions without full retraining~\cite{huang2024embedding}. \\

\textbf{RAG for recommendation: rapid progress, limited multimodal infrastructure.}
Recent work has explored RAG-style recommendation in several directions, including conversational RAG pipelines~\cite{qiu2025graph} and knowledge-grounded LLM recommenders that retrieve structured evidence (e.g., from knowledge graphs) before generation~\cite{wang2025knowledge,deldjoo2024recommendation}. However, these systems typically rely on task-specific corpora (dialogue logs, knowledge documents, or KG triples) and are often not directly reusable for \emph{multimodal} movie recommendation, where the retrieved evidence must integrate both \emph{textual} and \emph{visual} item signals and support evaluation
under standard MovieLens protocols. As a result, there remains a gap in \emph{open, auditable} benchmarks and toolkits that enable reproducible, extensible multimodal RAG experimentation on movies.

\noindent \textbf{Why video?}
The movie domain is an ideal yet challenging test-bed for multimodal RAG recommenders: items are naturally multimodal (trailers, visuals, audio, plot text), while metadata is often sparse, inconsistent,
or missing for long-tail titles. Effective systems must (i) fuse heterogeneous modalities, (ii) retrieve reliable candidates under sparse signals, and (iii) support generation and re-ranking without sacrificing reproducibility.

\begin{table*}[t!]
    \centering
    \caption{Comparative analysis of (A) multimodal recommendation resources/systems and (B) RAG-based recommenders.
    We separate \emph{multimodal CF-focused} toolkits/datasets from \emph{RAG-grounded} LLM recommenders, since the latter
    often require additional evidence sources (dialogue corpora, knowledge graphs, or external documents) that are not
    available in standard movie benchmarks.}
    \label{tbl:sample_table}
    \begin{adjustbox}{max width=\textwidth}
    \begin{tabular}{l c c c c l c c c c}
    \toprule
    \textbf{Name} & \textbf{Type} & \multicolumn{3}{c}{\textbf{Modality}} & \textbf{Fusion / Retrieval Evidence} &
    \textbf{CF} & \textbf{RAG} & \textbf{Multimodal RAG} & \textbf{Data/Code} \\
    \cmidrule(lr){3-5}
     &  & V & A & T &  &  &  &  &  \\
    \midrule
    \rowgray Ducho v2.0~\cite{ducho2} & Toolkit & \cmark & \cmark & \cmark & Sum/Mul/Concat/Mean & \cmark & \xmark & \xmark & \href{https://github.com/sisinflab/Ducho}{[code]} \\
    Ducho-meets-Elliot~\cite{ducho_elliot} & Toolkit & \cmark & \cmark & \cmark & Sum/Mul/Concat/Mean & \cmark & \xmark & \xmark & \href{https://github.com/sisinflab/Ducho-meets-Elliot}{[code]} \\
    \rowgray MMSSL~\cite{mmsl} & Method & \cmark & \cmark & \cmark & Modality-aware fusion & \cmark & \xmark & \xmark & \href{https://github.com/HKUDS/MMSSL}{[code]} \\
    MMRec~\cite{mmrec} & Method & \cmark & \cmark & \cmark & Concat & \cmark & \xmark & \xmark & \href{https://github.com/enoche/MMRec}{[code]} \\
    \rowgray MMRec-LLM~\cite{mmrec_llm} & Method & \cmark & \xmark & \cmark & Concat + LLM summarization & \cmark & \xmark & \xmark & \xmark \\
    MicroLens~\cite{microlens} & Toolkit & \cmark & \cmark & \cmark & Sum/Concat & \cmark & \xmark & \xmark & \href{https://github.com/westlake-repl/MicroLens}{[code]} \\
    \rowgray MMTF-14K~\cite{mmtf14k} & Dataset & \cmark & \cmark & \cmark & Mean/Median aggregation & \xmark & \xmark & \xmark & \href{https://zenodo.org/records/1225406}{[data]} \\
    YouTube-8M~\cite{youtube8m} & Dataset & \cmark & \xmark & \cmark & PCA features (video-level) & \xmark & \xmark & \xmark & \href{https://research.google.com/youtube8m/}{[data]} \\
    \midrule
    \rowgray Rec-GPT4V~\cite{recgpt} & Method-RAG RecSys & \cmark & \xmark & \cmark & VLM prompting (no open benchmark) & \cmark & \cmark & \cmark & \xmark \\
    G-CRS~\cite{qiu2025graph} & Method-RAG RecSys & \xmark & \xmark & \cmark & Graph / doc evidence retrieval & \xmark & \cmark & \xmark & (varies) \\
    \rowgray K-RagRec~\cite{wang2025knowledge} & Method-RAG RecSys & \xmark & \xmark & \cmark & Knowledge-graph grounded retrieval & \xmark & \cmark & \xmark & (varies) \\
    \midrule
    \rowgray \textbf{\RAGBench~(ours)} & Benchmark/Toolkit & \cmark & \xmark & \cmark & Concat/PCA/CCA + kNN retrieval & \cmark & \cmark & \cmark &
    \href{https://anonymous.4open.science/r/RAG-VisualRec-2866/}{[code]} \\
    \bottomrule
    \end{tabular}
    \end{adjustbox}
\end{table*}

\textbf{Gap in existing datasets and systems.}
Despite notable advances in multimodal recommendation, current resources remain limited for rigorous multimodal RAG evaluation (Table~\ref{tbl:sample_table}). Many datasets and toolkits focus on multimodal \ac{CF} with simple fusion but do not provide an explicit RAG loop with modular retrieval/augmentation/generation stages, nor do they provide protocols for auditing LLM-generated side information used for augmentation. In addition, many recent RAG-based recommenders require extra evidence sources (knowledge graphs, dialogue corpora, or external documents) that are not available in standard movie benchmarks, making direct experimental comparison non-trivial without substantial additional data construction.

In Table~\ref{tbl:sample_table}, the upper block primarily contains multimodal \ac{CF}-oriented datasets/toolkits and fusion-based recommenders: they are useful for learning multimodal representations, but they typically do not expose an end-to-end RAG protocol with explicit \emph{retrieval--augmentation--generation} stages. The middle block summarizes recent RAG-style recommenders, which usually ground LLM generation in external \emph{textual or structured} evidence (e.g., documents, dialogue corpora, or knowledge graphs). Such evidence sources are often task-specific and are not readily available in standard MovieLens-style evaluation setups, making direct comparison difficult without substantial additional data construction.

\RAGBench\ is designed to bridge this gap at the \emph{infrastructure} level. It releases an open movie benchmark that aligns user interactions with multimodal item evidence (trailer-derived visual features plus enriched textual signals), and it provides a modular pipeline (fusion, vector retrieval, optional prompting, and LLM re-ranking) together with a configuration-driven benchmarking harness for reproducible evaluation. Crucially, \RAGBench\ also supports \emph{LLM-eval} of synthetic side information via \texttt{LLMGenQC} (\S~\ref{sec:ragmovieaug}): each generated synopsis is accompanied by auditing scores and an accept/reject label covering semantic alignment to metadata anchors, internal consistency, safety, and basic sanity checks. These artifacts are released alongside the dataset to enable transparent filtering, benchmarking, and future extensions under the same protocol.

\paragraph{Evaluating LLM-generated side information.}
Surveys on recommendation with generative models and RAG~\cite{deldjoo2024recommendation,gao2023retrieval} note that LLMs can enrich sparse item metadata (e.g., cold-start items with missing or incomplete side information), but also emphasize that \emph{evaluating} synthetic content is non-trivial in the absence of ground-truth labels. Accordingly, we audit all LLM-generated item descriptions using a critic suite (\S~\ref{sec:ragmovieaug}) and release the resulting critic scores and a binary accept/reject label alongside the synthetic synopses for reproducible downstream use.

\textbf{Positioning and comparative context.}
Table~\ref{tbl:sample_table} contrasts representative multimodal resources (mainly \ac{CF}-oriented) and recent RAG-based recommenders (often text- or KG-grounded) against \RAGBench. We position \RAGBench\ as \emph{research infrastructure}---an open, auditable benchmark and modular toolkit for multimodal movie RAG---rather than a new leaderboard-optimized model. The released framework provides reusable fusion modules, a configurable RAG loop, and a benchmarking harness that supports controlled re-implementation and extension of prior methods under a transparent setup.

\noindent
\textbf{RAG-VisualRec at a glance.}\footnote{\url{https://github.com/RecSys-lab/RAG-VisualRec}}
To make the above positioning concrete, we release \RAGBench\ (RAG-VisualRec) as an \emph{open, auditable} multimodal benchmark and toolkit for \emph{multimodal RAG} in movie recommendation. The aim is not to introduce a new leaderboard-optimized architecture, but to provide missing \emph{research infrastructure} that (i) aligns multimodal signals under a unified retrieval-and-generation pipeline, (ii) makes evaluation protocols explicit (including two-stage cut-offs and beyond-accuracy metrics), and (iii) exposes \emph{auditing metadata} for LLM-generated side information so that future work can quantify and control risks such as semantic drift, contradictions, or unsafe generations.

\begin{itemize}[leftmargin=1.1em]
    \item \textbf{Dataset layer (multimodal alignment + augmentation).}
    We align 9{,}724 titles from \textsc{MovieLens-Latest} with the 3{,}751 trailers available in the public \textsc{MMTF-14K} collection, enabling joint study of interaction data and item-side multimodal content. Each trailer is shipped with multiple pre-extracted feature families (e.g., CNN-based frame embeddings and aesthetic-style descriptors) under both \textsc{Average} and \textsc{Median} aggregation schemes; our experiments use the \emph{CNN / Average} variant, while other representations are enabled via a single configuration flag. The package also includes the block-level and i-vector \emph{audio} descriptors provided by \textsc{MMTF-14K} (not exploited in the current study but retained as a ready-to-use modality for follow-up work). To mitigate sparse metadata (e.g., only title/genres/tags) and strengthen cold-start coverage on the item side, we additionally provide GPT-4o--generated plot synopses and rationales that can be embedded and retrieved as textual evidence.

    \item \textbf{Toolkit layer (modular multimodal RAG + baselines).}
    We provide a typed, open-source library that implements the end-to-end pipeline components needed to reproduce and extend multimodal RAG recommenders: embedding extraction hooks, early- and mid-fusion operators (Concat/PCA/CCA), retrieval over vector stores (cosine kNN), optional augmentation prompts, strong \ac{CF} baselines, and an LLM-based selection/re-ranking stage. The design is intentionally modular: the same dataset layer can be paired with different backbones (text encoders, visual encoders) and different fusion operators, enabling controlled comparisons.

    \item \textbf{Benchmark harness (explicit protocol + reproducibility).}
    We release a declarative configuration schema that regenerates experiments end-to-end, including all metrics, tables, and plots. Importantly, the harness makes the two-stage evaluation protocol explicit: retrieval is evaluated at cut-off $N$ (Recall@N / nDCG@N for the candidate set), whereas the final recommendation list is evaluated at cut-off $K$ (Recall@K / nDCG@K, with $K{=}10$ fixed throughout). The full system supports 12 accuracy and 5 beyond-accuracy indicators (e.g., coverage, novelty, long-tail share, diversity); due to space constraints, this paper reports a focused subset (Recall, nDCG, Novelty, LongTailFrac, and catalogue coverage).
\end{itemize}

 \begin{figure}[!t]
    \centering
     \includegraphics[width=\textwidth]{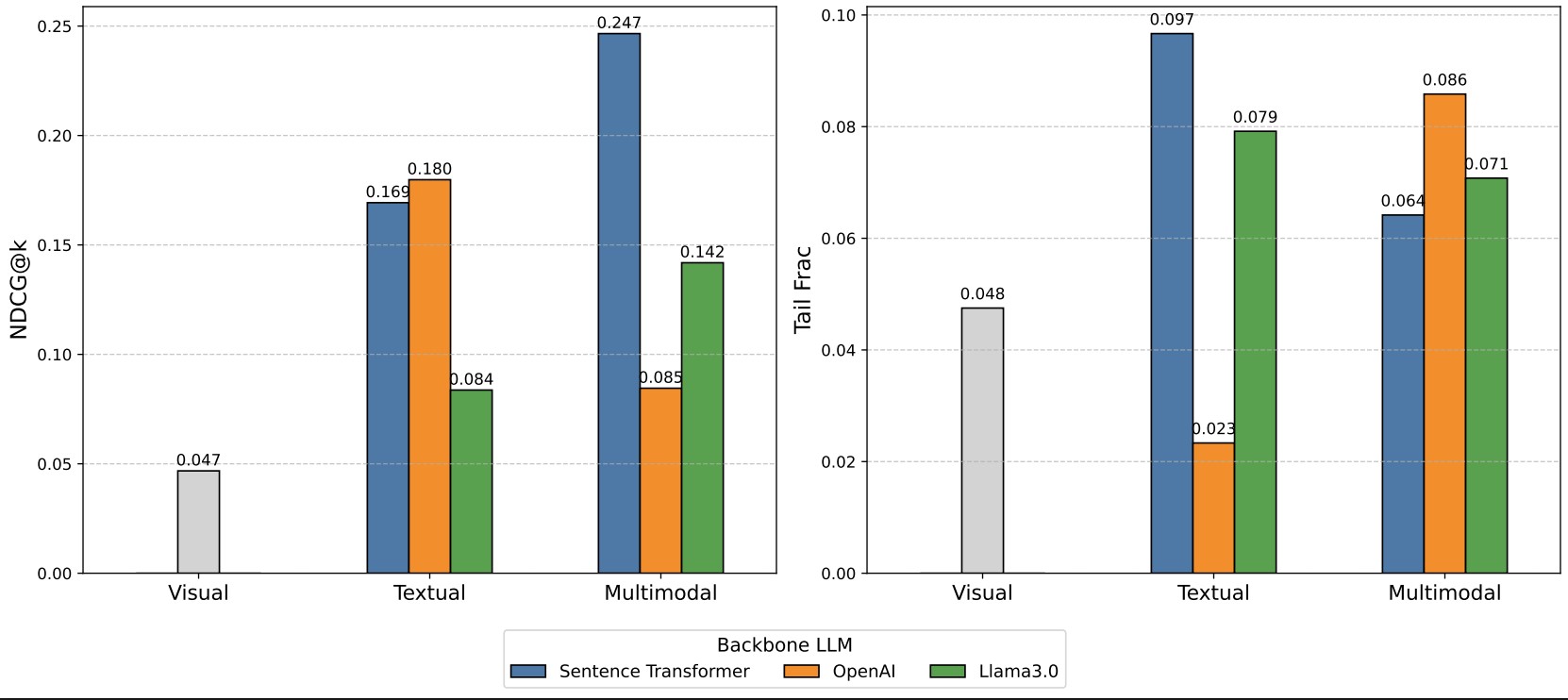}
     \caption{Overall performance overview.}
     \label{fig:results}
 \end{figure}

\noindent\textbf{Contributions.}
We position this work as an infrastructural contribution consolidating data, tooling, and evaluation for multimodal RAG-based
recommendation. Concretely, the paper provides:

\begin{enumerate}[label=\arabic*.,leftmargin=2.3em]
    \item \textbf{End-to-end multimodal RAG pipeline.}
    An executable pipeline that combines trailer-derived visual embeddings with text embeddings from modern language models and
    supports three fusion strategies: Concat, PCA, and CCA.

    \item \textbf{Open multimodal resource for movies.}
    An aligned movie--trailer corpus with multiple visual descriptor variants and audio descriptors (available for future work),
    plus LLM-enriched textual descriptions to support retrieval-augmented generation under sparse-metadata conditions.

    \item \textbf{Auditable LLM augmentation via \texttt{LLMGenQC}.}
    A critic-based quality-control component that assigns each generated synopsis a suite of scores (semantic alignment to
    metadata anchors, contradiction/consistency checks, toxicity/safety checks, and sanity signals) and a final accept/reject
    label. All critic scores and labels are released alongside the synthetic text to support transparent filtering and future
    benchmarking; the audit is validated via a toy study reported in \textbf{RQ5}.

    \item \textbf{Reproducible evaluation suite (accuracy + beyond-accuracy).}
    A unified evaluation protocol with formal metric definitions and a configuration-driven harness that reproduces results
    end-to-end.

    \item \textbf{Empirical validation and ablations.}
    Experiments showing consistent gains from CCA-based fusion over unimodal and Concat baselines, and improvements from GPT-4o
    re-ranking in nDCG@10; parameter sensitivity and ablation studies (retrieval depth, fusion settings, and user-embedding
    construction) are reported in \textbf{RQ6}.
\end{enumerate}

\noindent\textbf{Experimental scope.}
Figure~\ref{fig:results} previews key results by comparing key metrics (nDCG\@10 and tail fraction) across different text backbones (Sentence Transformers, OpenAI, Llama3) under three setups: visual-only, text-only, and multimodal fusion. The fused multimodal approach consistently outperforms unimodal variants in nDCG\@10, particularly for Sentence Transformers and Llama3, and typically enhances catalogue coverage compared to both visual-only and text-only baselines. Our experiments also demonstrate that Canonical Correlation Analysis (CCA)--based multimodal fusion consistently surpasses unimodal and simple-concat baselines, delivering up to \textbf{27\%} higher Recall@10 and almost doubling catalog coverage.

In summary, \RAGBench\ provides a reproducible test-bed for multimodal RAG research in movie recommendation and releases
data, code, configuration files, and auditing artifacts to facilitate controlled comparisons and future extensions.

\section{Toolkit Design and Visual-RAG Pipeline}
\label{sec:visualrag}

\subsection{System Overview: Motivation and Researcher Experience}

\begin{figure}[t!]
     \centering
     \includegraphics[width=1.0\textwidth]{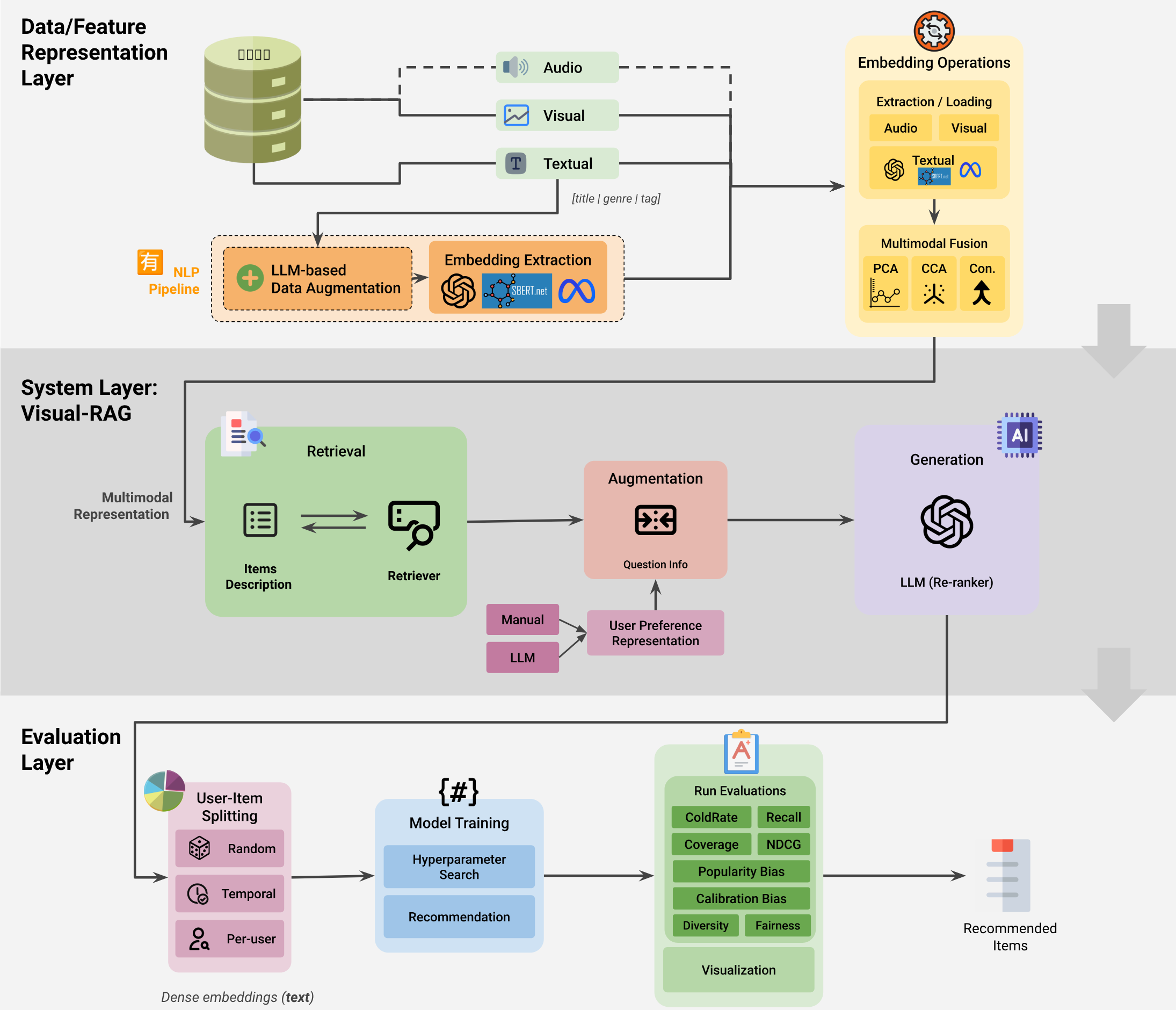}
     \caption{Visual-RAG: Modular architecture supporting data ingestion, multimodal embedding, fusion, LLM-based data augmentation, \textbf{critic-based quality auditing of synthetic texts (\texttt{LLMGenQC} )}, retrieval, LLM re-ranking, and robust evaluation.}
     \label{fig:pipeline2}
\end{figure}

The \textbf{Visual-RAG pipeline} is designed as a transparent, modular, and extensible resource for rigorous multimodal recommendation research. Its primary goal is to bridge the gap between theoretical advances (e.g., fusion techniques, textual data augmentation, multi-modal retrieval, and augmented generation) and practical, reproducible workflows that any researcher can adapt or extend. Figure~\ref{fig:pipeline2} visualizes the pipeline as a series of interconnected, parameterized blocks, enabling rapid experimentation and seamless integration of new modalities, textual data augmentation, or evaluation metrics.

\textbf{Researcher Scenario:} Suppose a movie recommendation researcher wants to benchmark the impact of multimodal fusion (textual + visual) under cold-start conditions, simulate the impact of data augmentation on popular movies, and reproduce all metrics with a single config file—Visual-RAG enables all this without any code edits.

\subsection{Step-by-Step Pipeline Walkthrough}

\begin{enumerate}[leftmargin=2.5em,label=\ding{\arabic*},itemsep=1.5em]

\item \textbf{Data Preparation and Ingestion} \\
Data ingestion begins with MovieLens-Latest-Small or MovieLens-1M, loaded via \texttt{pandas.DataFrame} adapters for uniformity. All user-item interactions are cleaned, integer-indexed, and merged with available metadata (\textcolor{teal}{title}, \textcolor{teal}{genres}, \textcolor{teal}{tags}). The pipeline handles missing metadata gracefully, a crucial requirement for studying real-world cold-start scenarios or incomplete catalogs.

\begin{tcolorbox}[colback=yellow!8!white, colframe=yellow!80!black, sharp corners=south, boxrule=0.4pt, left=1mm, right=1mm, top=0.5mm, bottom=0.5mm]
\textbf{Worked Example: Data Enrichment}\\
\textcolor{magenta}{\textbf{Input:}} Title = \textcolor{teal}{Nixon (1995)}, Genres = \textcolor{teal}{Drama, Biography}, Description = \textcolor{gray}{(missing)} \\
\textcolor{magenta}{\textbf{Output after LLM Augmentation:}}
\textcolor{blue}{``Nixon (1995) explores the troubled psyche and political career of America's 37th president, delving into both his strategic brilliance and the moral compromises that shaped his legacy. Directed by Oliver Stone, the film highlights Nixon's complex relationships with political allies and adversaries, offering a gripping portrayal of power and vulnerability.''}
\end{tcolorbox}

To enable targeted experiments, a \textcolor{purple}{popularity annotator} stratifies all items into \textbf{head} (top 10\%), \textbf{mid-tail} (next 40\%), or \textbf{long-tail} (bottom 50\%) based on their frequency in the data. This annotation is vital for controlled cold-start evaluation.

All such LLM-generated enrichments are subsequently passed through the 	exttt{LLMGenQC} critic suite
(Section~\ref{sec:ragmovieaug}), which logs SBERT-based relevance, perplexity, toxicity, and NLI-based
contradiction scores for each synopsis. The resulting quality scores and pass/fail flags are stored
alongside the augmented descriptions, so that downstream experiments can either filter or reweight
synthetic item texts according to their desired safety and fidelity requirements.

\item \textbf{Multimodal Embedding Extraction} \\
      This stage generates unified item representations from multiple modalities:
      \begin{itemize}[leftmargin=1.5em,topsep=0.1pt]
          \item[(a)] \textbf{Textual:} Item descriptions are embedded using one of three back-ends: OpenAI-Ada, SentenceTransformer-MiniLM, or \emph{LLaMA-2/3} (via HuggingFace). This enables evaluation under both proprietary and open-source large language model (LLM) encoders. The text-to-vector mapping yields $d_{\text{text}}$-dimensional outputs.
          \item[(b)] \textbf{Visual:} Frame-level visual features are extracted using ResNet-50, with both mean and median pooling across sampled keyframes to produce robust $d_{\text{vis}} = 2048$-dimensional vectors per item.
          \item[(c)] \textbf{Audio (optional):} Four Mel-Frequency Cepstral Coefficient (MFCC) variants (log, correlation, delta, spectral) are computed, each resulting in $d_{\text{aud}}=128$-dimensional vectors, providing complementary semantic cues for music or video datasets.
          \item[(d)] \textbf{Fusion:} Multimodal fusion is implemented using four stateless operators—\textsc{concat}, \textsc{pca}$_{128}$, \textsc{cca}$_{64}$, and \textsc{avg}—which project concatenated features into a joint latent space $d^\star$. These operators allow for controlled ablation of cross-modal information and support both early and late fusion paradigms.
      \end{itemize}

\textbf{Fusion Operators:} These modality-specific embeddings are combined using a choice of stateless operators, set via a configuration flag:
\begin{itemize}[leftmargin=1.7em, topsep=0.1em]
\item \textbf{Concatenation} (\texttt{concat}): Directly stacks all vectors.
\item \textbf{PCA} (\texttt{pca\_128}): Projects concatenated vectors onto the first 128 principal components.
\item \textbf{CCA} (\texttt{cca\_64}): Aligns textual and visual spaces, maximizing cross-modal correlation.
\item \textbf{Averaging} (\texttt{avg}): Arithmetic mean for sanity checking.
\end{itemize}
This setup allows for plug-and-play ablation—simply change the fusion method in config and rerun to study its impact.

\begin{tcolorbox}[colback=cyan!5!white, colframe=cyan!70!black, sharp corners=south, boxrule=0.4pt, left=1mm, right=1mm, top=0.5mm, bottom=0.5mm]
    \textbf{Example: Fusion in Practice}\\
    \textcolor{blue}{Text embedding} of \textcolor{teal}{Nixon (1995)}: $[\cdots, -0.31, 0.54, 1.02, ...]$ \\
    \textcolor{blue}{Visual embedding} from trailer frames: $[\cdots, 0.11, -0.22, 0.91, ...]$ \\
    \textcolor{purple}{CCA fusion} produces a joint 64-d vector: $[\cdots, 0.44, 0.08, -0.32, ...]$
\end{tcolorbox}

\item \textbf{Embedding Swap and Re-embedding}
When item side information is altered (e.g., via adversarial augmentation), the pipeline re-embeds those items using the original encoder settings. This ensures that all subsequent retrieval and ranking steps are based on the current (potentially poisoned) representations—enabling true end-to-end robustness studies.

\item \textbf{User Embedding Construction}
User representations are constructed with three interchangeable strategies:
\begin{itemize}[leftmargin=1.8em, topsep=0.1em]
\item \textbf{Random}: Assigns a random vector as a baseline (for sanity checks).
\item \textbf{Average}: Averages the fused embeddings of all high-rated items for the user.
\item \textbf{Temporal}: Applies logistic decay to prioritize recent interactions.
\end{itemize}
All user vectors are mapped to the same fused space as items, allowing direct cosine-based retrieval.

\begin{tcolorbox}[colback=lime!7!white, colframe=lime!80!black, sharp corners=south, boxrule=0.4pt, left=1mm, right=1mm, top=0.5mm, bottom=0.5mm]
\textbf{Example: Temporal Embedding}\\
Suppose User 42 has rated \textcolor{teal}{Nixon (1995)}, \textcolor{teal}{The Post}, and \textcolor{teal}{Frost/Nixon}. Each rating is timestamped. Temporal embedding weights the most recent (e.g., \textcolor{teal}{The Post}) more heavily in the user's fused profile.
\end{tcolorbox}

\item \textbf{Candidate Retrieval}
Given a user embedding, the pipeline constructs a cosine-based $k$-NN index over all item vectors, efficiently returning the top-$N$ most similar items for each user. This step is sensitive to all upstream choices—embedding models, fusion, and user profile method—enabling direct comparison of retrieval quality under varied research conditions.

\item \textbf{Profile Augmentation and LLM Prompting} \\
For LLM-based re-ranking, the system composes a structured \textcolor{purple}{JSON} user profile, containing favorite genres, most-used tags, top items, and a free-form taste synopsis. Profiles can be:
\begin{itemize}[leftmargin=1.7em, topsep=0.1em]
\item \textbf{Manual:} Derived by code from historical data (e.g., top genres, recent items).
\item \textbf{LLM-based:} Synthesized from user history via LLM prompt (e.g., ``This user enjoys political dramas and historical biographies...'').
\end{itemize}

The final prompt to the LLM includes (a) the user profile, (b) candidate items with metadata, and (c) strict task instructions (e.g., ``Return the top 10 recommended item IDs as a JSON array'').

\begin{tcolorbox}[colback=purple!7!white, colframe=purple!70!black, sharp corners=south, boxrule=0.4pt, left=1mm, right=1mm, top=0.5mm, bottom=0.5mm]
\textbf{Example: User Profile Prompt}\\
\textcolor{gray}{\texttt{USER PROFILE:}}\\
\textcolor{blue}{\texttt{Genres: ["Drama", "Biography"], Top items: ["Nixon (1995)", "The Post"], Taste: "Prefers political dramas exploring real historical events."}}\\
\textcolor{gray}{\texttt{TASK: Given this profile and the following candidate movies, return your top-10 recommendations as a JSON list of item IDs.}}
\end{tcolorbox}

\item \textbf{LLM Generation Head and Re-ranking} \\
The LLM-based generation head accepts the profile and candidates, then outputs a ranked list of recommendations. Two privacy regimes are supported:
\begin{itemize}[leftmargin=1.7em, topsep=0.1em]
\item \textbf{ID-only}: Returns only recommended item IDs (for privacy, e.g., production use).
\item \textbf{Explainable}: Returns both IDs and explanations/chain-of-thought rationales (for interpretability research).
\end{itemize}
A robust JSON parser extracts outputs; failures fall back to the kNN list, ensuring robustness.

\item \textbf{Evaluation and Logging} \\
Each experiment is evaluated with a comprehensive suite of metrics:
\begin{itemize}[leftmargin=1.7em, topsep=0.1em]
\item \textbf{Accuracy}: Recall@K, nDCG@K, MAP, MRR, etc.
\item \textbf{Beyond-Accuracy}: Coverage, novelty, diversity, long-tail fraction.
\item \textbf{Fairness/Robustness}: Cold-start rate, exposure.
\end{itemize}
Metrics are computed per-user, averaged across test splits, and exported as CSV/Parquet. Intermediate artefacts (embeddings, poisoned texts, candidate lists) are checkpointed, ensuring full reproducibility. Optionally, all outputs can be pushed to a public repository for open auditing.

\end{enumerate}

\subsection{Configuration Schema}

Each stage of the pipeline is fully configurable through a centralized parameter block in the Colab notebook. Key hyperparameters and toggles include:
\begin{itemize}[leftmargin=1.25em,itemsep=3pt]
  \item \textbf{Dataset}: \texttt{ml\_latest\_small}\,|\,\texttt{lfm360k}
  \item \textbf{Embeddings}: \texttt{textual | visual | audio | fused\_\{concat,pca,cca,avg\}}
  \item \textbf{LLM Model}: \texttt{openai | sentence\_transformer | llama3}
  \item \textbf{User Vector}: \texttt{random | average | temporal}
  \item \textbf{Retrieval}: $N$ (default 50)
  \item \textbf{Recommendation}: $K$ (default 10), \texttt{explainable?}
  \item \textbf{Runtime}: \texttt{use\_gpu}, \texttt{seed}, \texttt{batch\_size}
\end{itemize}

Default values are chosen to reproduce the benchmarks and ablation studies exactly reported in Section~\ref{sec:benchmark_experiments}. Importantly, changes to any flag or hyperparameter require no code edits, supporting rapid experimentation and extensibility.
The pipeline is fully controlled by a single config block that specifies the dataset, embedding/fusion methods, user profile mode, retrieval depth, evaluation settings, and runtime (GPU/CPU, batch size, random seed). Modifying any experimental setting requires only a config edit, not a code change.

\noindent
\textbf{Implementation Footprint:} On a standard Google Colab A100 (40 GB), a full run (MovieLens-1M, all fusion variants) completes in $\sim$3 hours, with peak memory under 11 GB. This efficiency enables large-scale or multi-domain experiments even on modest hardware.

\subsection{Practical Implications and Researcher Perspective}

\noindent
\textbf{Reproducibility and Transparency:} All experiments are deterministic (seeded), versioned, and documented. CI jobs and open-source scripts enable reviewers to audit every table or figure. Rapid switching between baselines or fusion methods facilitates systematic ablation and robustness research.

\noindent
\textbf{Research Impact:} The Visual-RAG pipeline empowers researchers to:
\begin{itemize}[leftmargin=1.2em]
\item Benchmark the effect of multimodal fusion under various data sparsity and cold-start conditions.
\item Study trade-offs between explainability, privacy, and accuracy.
\end{itemize}

\noindent
\textbf{Summary:} Through modular design, extensive configuration, worked examples, and fully open infrastructure, Visual-RAG offers a transparent, extensible, and auditable testbed for next-generation recommender system research.

\section{Formalism}
\label{sec:formal}
In the following section, we formalize the proposed multimodal framework leveraging RAG for movie recommendation. 
\subsubsection*{Item Representations and Multimodal Fusion.}
\label{subsec:item_rep}

Let $\mathcal{U}$ denote the set of users and $\mathcal{I}$ denote the set of items (i.e., movies).
Each item $i \in \mathcal{I}$ can be represented by two primary embeddings:
\begin{itemize}
    \item \textbf{Textual embedding}: $\mathbf{x}_i^{(\mathrm{txt})} \in \mathbb{R}^{d_{\mathrm{txt}}}$, extracted from a \acl{LLM} (e.g., GPT, ST, or Llama).
    \item \textbf{Visual embedding}: $\mathbf{x}_i^{(\mathrm{vis})} \in \mathbb{R}^{d_{\mathrm{vis}}}$, extracted from trailer frames or short promotional clips.
\end{itemize}
If available, audio embeddings $\mathbf{x}_i^{(\mathrm{aud})} \in \mathbb{R}^{d_{\mathrm{aud}}}$ may also be included.
A \emph{fusion function} $F(\cdot)$ combines these embeddings into a single \emph{multi-modal} vector $\mathbf{z}_i \in \mathbb{R}^{d_{z}}$.
\emph{Common approaches} include:
\begin{align}
    & \mathbf{z}_i \;=\; \mathrm{PCA}\Bigl(\mathbf{x}_i^{(\mathrm{txt})} \,\Vert\, \mathbf{x}_i^{(\mathrm{vis})}\Bigr),
    \\[5pt]
    & \bigl(\mathbf{z}_i^{(\mathrm{txt})}, \mathbf{z}_i^{(\mathrm{vis})}\bigr)
        \;=\;\mathrm{CCA}\bigl(\mathbf{x}_i^{(\mathrm{txt})}, \mathbf{x}_i^{(\mathrm{vis})}\bigr),
\end{align}
where $\Vert$ denotes concatenation. 
\emph{With PCA}, we first concatenate text and visual embeddings, then project to $d_z$ dimensions. 
\emph{With CCA}, we learn $\mathbf{z}_i^{(\mathrm{txt})}$ and $\mathbf{z}_i^{(\mathrm{vis})}$ as canonical components in a shared subspace.
We \emph{concatenate} these canonical components into a single vector:
\[
  \mathbf{z}_i \;=\; \left[\mathbf{z}_i^{(\mathrm{txt})} \,\Vert\, \mathbf{z}_i^{(\mathrm{vis})}\right]
\]

\noindent\textbf{Why CCA might outperform PCA (intuition).}
PCA performs a variance-preserving projection in the \emph{concatenated} space $(x_i^{txt}\,\|\,x_i^{vis})$,
so the resulting components are dominated by directions of \emph{high marginal variance} (often modality-specific),
which may not correspond to shared semantics across modalities. In contrast, CCA learns paired projections
$(W_{txt}, W_{vis})$ that maximize the \emph{cross-modal correlation} between $W_{txt}^\top x_i^{txt}$ and
$W_{vis}^\top x_i^{vis}$, explicitly emphasizing aligned factors (e.g., genre-related text cues that co-vary with
visual style cues) while down-weighting uncorrelated, modality-specific variation. This correlation objective
often yields a fused representation with stronger cross-modal synergy, which is consistent with the empirical
improvements observed in Section~\ref{sec:rq2}.

Both approaches leverage complementary cues (semantic text vs.\ visual aesthetics), thereby enabling multimodal fusion, which can result in enhanced movie recommendation performance.
Each user $u \in \mathcal{U}$ has historically rated or interacted with a subset of items. In a content-driven embedding approach, we compute the user embedding $\mathbf{u}_{u}$ by aggregating item embeddings. The \emph{average} user embedding is:
\[
   \mathbf{p}_{u}^{(\text{avg})} \;=\;
   \frac{1}{\bigl|\mathcal{I}_{u}^{+}\bigr|}\,\sum_{i \in \mathcal{I}_{u}^{+}}\,\mathbf{z}_{i},
\]
where $\mathcal{I}_{u}^{+}$ is the set of items that $u$ rated above a threshold (e.g.\ 4 stars). The \emph{temporal} embedding instead weights items by recency:
\[
   \mathbf{p}_{u}^{(\text{temp})} \;=\;
   \frac{\sum_{i \in \mathcal{I}_{u}^{+}}\,\bigl(\mathbf{z}_i \cdot w_{u,i}\bigr)}{\sum_{i \in \mathcal{I}_{u}^{+}}\,w_{u,i}},
   \quad
   w_{u,i} \;=\; \frac{1}{\,1+\exp\!\bigl[-\alpha\,(\mathrm{time}(i)-\bar{t}_u)\bigr]},
\]
where $\bar{t}_u$ denotes the average timestamp for user $u$, and $\alpha$ is a smoothing hyperparameter.

\begin{tcolorbox}[colback=lightgray!15, colframe=gray!50, boxrule=0.5pt, sharp corners]
\textbf{Note.} We experimented with both user profile representations, $\mathbf{p}_{u}^{(\text{avg})}$ and $\mathbf{p}_{u}^{(\text{temp})}$, ultimately selecting the latter for its superior end-to-end performance, excluding the former for {space considerations}. Henceforth, in the following we denote $\mathbf{p}_{u} = \mathbf{p}_{u}^{(\text{temp})}$. The terms \textit{user profile embedding} and \textit{user embedding} are used interchangeably to refer to this representation.
\end{tcolorbox}

\begin{figure}[t]
    \centering
    \begin{subfigure}[t]{.22\columnwidth}
        \centering
        \includegraphics[width=\linewidth, trim=5 5 802 15, clip]{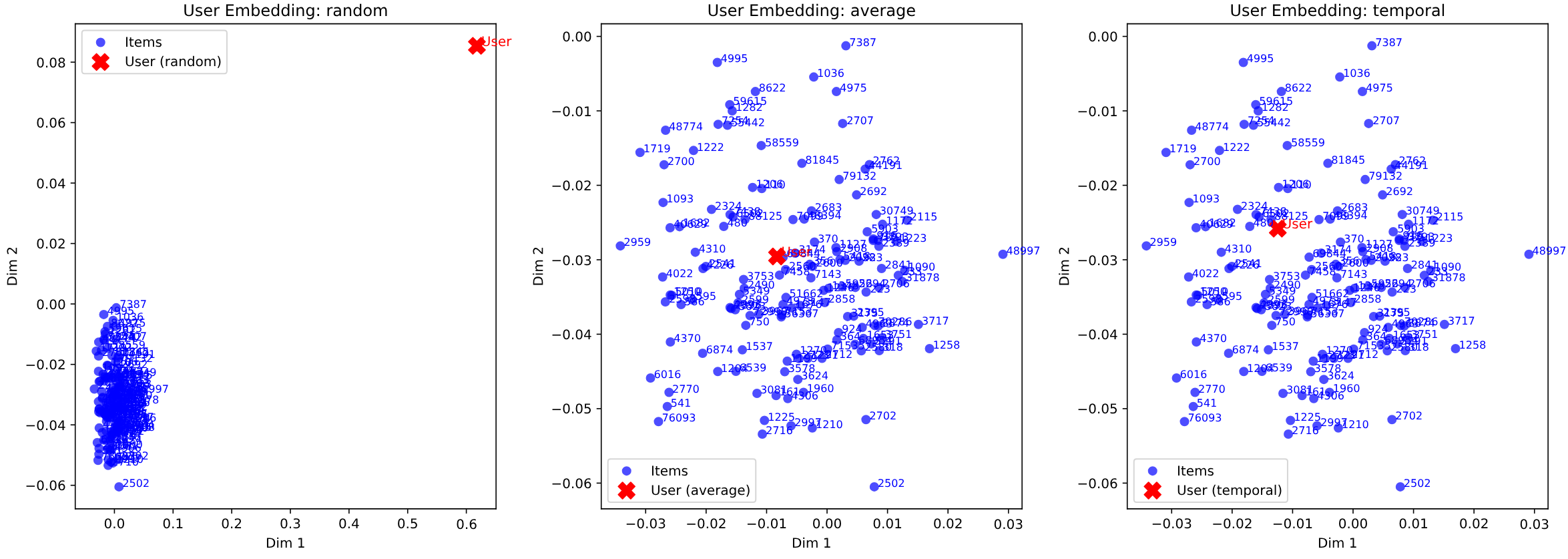}
        \caption{\scriptsize{Random}}
    \end{subfigure}
    \hspace{0.25mm}
    \begin{subfigure}[t]{.232\columnwidth}
        \centering
        \includegraphics[width=\linewidth, trim=792 7 0 15, clip]{textual.png}
        \caption{\scriptsize{Textual}}
    \end{subfigure}
    \hspace{0.25mm}
    \begin{subfigure}[t]{.228\columnwidth}
        \centering
        \includegraphics[width=\linewidth, trim=762 7 0 20, clip]{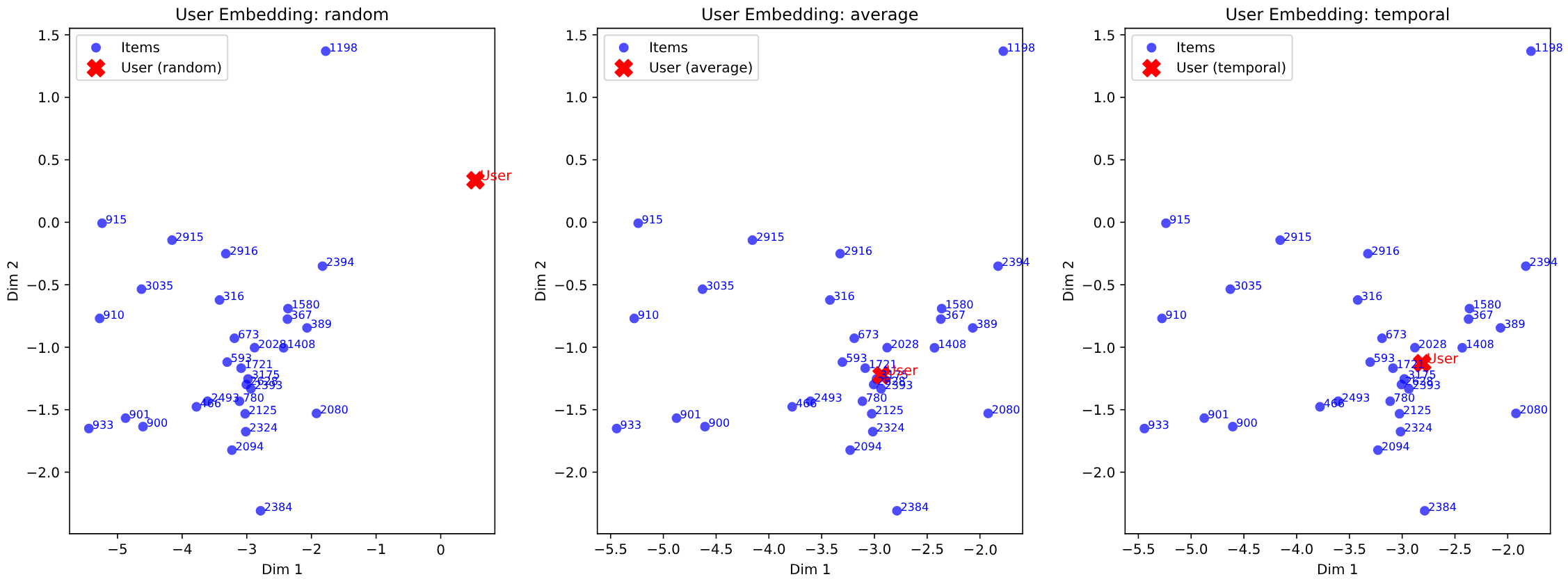}
        \caption{\scriptsize{Visual}}
    \end{subfigure}
    \hspace{0.25mm}
    \begin{subfigure}[t]{.23\columnwidth}
        \centering
        \includegraphics[width=\linewidth, trim=760 7 0 20, clip]{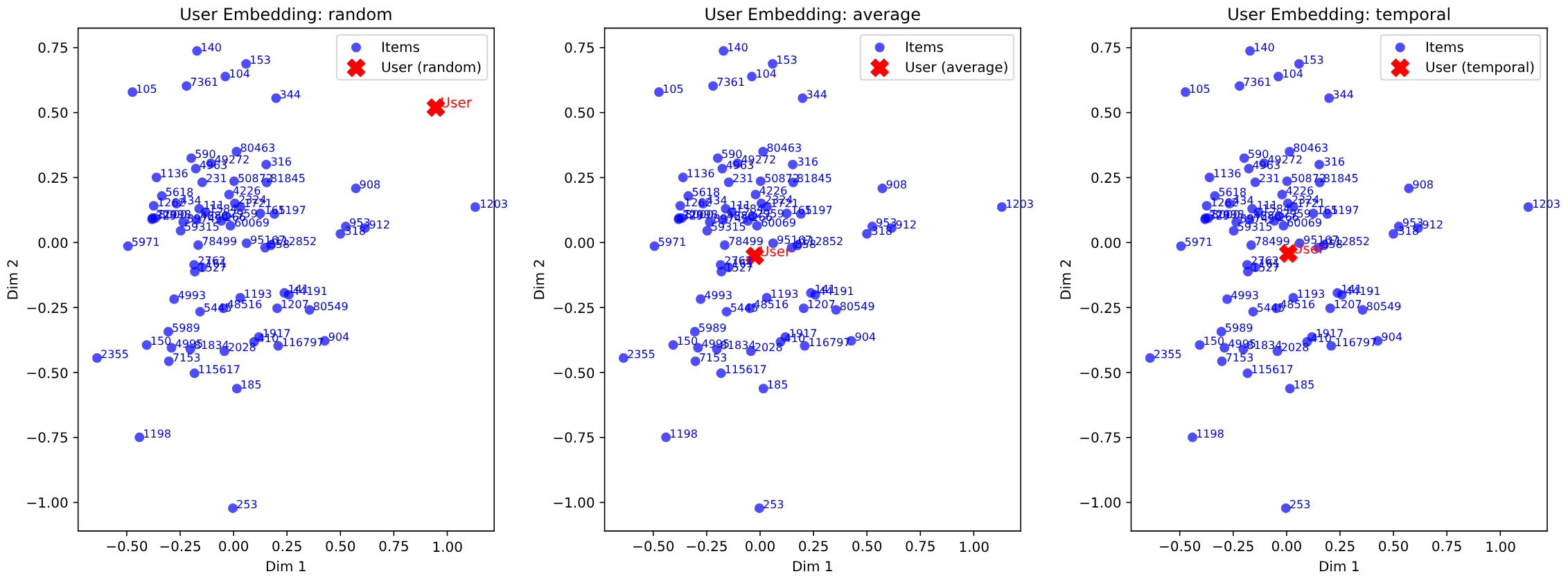}
        \caption{\scriptsize{Fused}}
    \end{subfigure}
\caption{t-SNE projection of item and user embeddings using the Sentence Transformer (ST) LLM backbone. Multimodal embeddings are obtained via CCA.}    \label{fig:embeddings}
\end{figure}

\noindent
\textbf{Visualization of t-SNE Projection of Embeddings.} As a demonstration, Figure~\ref{fig:embeddings} visualizes user embeddings (red stars) 
and item embeddings (blue points), in a shared 2D space, obtained via t-SNE projection of the embedding under investigation. investigation) using t-SNE. 
The four subplots represent (a) random user vectors, (b) textual, (c) visual, and (d) fused multimodal embeddings. 
In (a), the random user embedding is non-informative, clustering far from items. 
By contrast, (b), (c), and (d) employ the temporal aggregation strategies 
introduced in Section~\ref{subsec:item_rep}, producing more coherent user--item 
alignment. This demonstrates how textual, visual, and fused representations 
capture meaningful movie features, potentially enhancing recommendation performance.\footnote{We prioritize temporal user profiles for better recommendations, though T-SNE showed similar patterns for average embeddings.}

\subsubsection*{Retrieval-Augmented Generation (RAG) Stage}
\noindent \textbf{Retrieval Step.}
For recommendation with LLM-based generation, we first \emph{retrieve} top-$N$ items most relevant to user $u$.
In order to achieve this, based on the user profile embedding $\mathbf{p}_u$, we compute similarity $ s(u,i) \;=\; \cos\bigl(\mathbf{p}_u, \mathbf{z}_i\bigr)$ to produce final top-$k$ recommendation items, where $k \ll N$. We examine the impact of retrieval scope $N$ on enhancing the final top-$k$ performance in our experimental research questions (cf. \hl{\hyperref[sec:rq4]{RQ4}}).

\noindent \textbf{LLM-Based Augmentation.}
During augmentation, the LLM (e.g., GPT-4) takes as input
\begin{enumerate}[label=(\alph*)]
    \item A structured \emph{user profile} (genres, tags, top items),
    \item The \emph{candidate top-$N$ items} from retrieval, along with their metadata 
\end{enumerate}
into the form of a structured prompt. The LLM is then instructed to re-rank these $N$ candidates, generating a final list of $k$ recommended items, $k \le N$, plus optional textual rationales. We explore two methods to build or to generate this structured user profile:

\begin{enumerate}[label=(\roman*)]
    \item \textbf{Manual:} We compute a structured user profile (favorite genres, top-rated items, etc.) from the training data. A simple textual summary is created with a rule-based or heuristic script (i.e., “manual”).
    \item \textbf{LLM-based:} We feed the user’s historical items and preferences into a large language model (e.g., GPT-4) to generate a more free-form textual description of that user’s tastes.
\end{enumerate}

In our experimental research questions (RQs), we explore the impact of both augmentation strategies (cf. Sec. \ref{sec:benchmark_experiments}): 
\[
    \underbrace{\text{Manual Augmentation}}_{\text{structured template}} 
    \quad\text{vs.}\quad
    \underbrace{\text{LLM-based Augmentation}}_{\text{Free-Text}}
\]

\subsubsection*{Generation Step.}
\noindent In a typical \emph{two-stage} recommendation, the \emph{first stage (retrieval)} picks a broader set of candidates for efficiency, while the \emph{second stage (re-ranking)} refines the list using deeper signals (e.g., user textual profile). Formally:
\begin{equation*}
    \mathrm{Retrieve}_u(\mathbf{p}_u)\;=\;\underset{i \in \mathcal{I}}{\mathrm{argsort}_{i}}\;s\bigl(\mathbf{p}_u,\mathbf{z}_i\bigr)\quad[:k],
\end{equation*}
\begin{equation*}
    \mathrm{ReRank}_u\Bigl(\widetilde{\mathcal{C}}_{u}, \mathrm{Profile}_u\Bigr)\;=\;\mathrm{LLM}\Bigl(\widetilde{\mathcal{C}}_{u} \,\vert\, \mathrm{Profile}_u\Bigr),
\end{equation*}
where $\widetilde{\mathcal{C}}_{u}$ is the top-$N$ set from retrieval, and $\mathrm{Profile}_u$ is a textual or structured representation summarizing user $u$ preferences (genres, prior ratings, \textit{etc.}). This two-stage framework (\emph{retrieval} and \emph{LLM-based re-ranking/generation}) flexibly integrates multimodal embeddings (visual, textual, audio) into personalized recommendation tasks. Retrieval outputs (\(\widetilde{\mathcal{C}}_{u}\)) combined with user preference profiles (\(\mathrm{Profile}_u\)) naturally facilitate both structured recommendations and narrative generations.

\section{LLMGenQC. Critic-Based Quality Control for LLM-Generated Movie Metadata}
\label{sec:ragmovieaug}

\subsection{Motivation and goal.}
The MovieLens--Latest variant used in our main experiments is enriched with LLM-generated plot synopses,
which substantially increase item-side textual coverage. At the same time, synthetic metadata raises a
fundamental evaluation challenge: \emph{there is no ground truth for generated text}, hence quality cannot be
assessed by standard supervised metrics. In particular, LLM-generated synopses may (i) drift away from the
available metadata (semantic hallucination), (ii) contain internal inconsistencies or statements that
contradict the anchor information, or (iii) introduce unsafe content such as toxicity.
Rather than re-running the full recommendation pipeline with and without evaluation, we introduce
\texttt{LLMGenQC} as a self-contained \emph{toy study}: a transparent and
reproducible \emph{quality audit} of LLM-generated movie descriptions before downstream use.

\subsection{Evaluation Protocol.}
We operationalize quality assessment via an \emph{LLM-as-judge} (model-as-critic) approach: each generated
synopsis is scored by a suite of pretrained critic models that approximate complementary quality dimensions.
This choice is motivated by scalability and reproducibility: unlike human evaluation, critic models can be
run cheaply at scale, produce consistent scores across runs, and yield an auditable record of why a text was
kept or rejected. Importantly, we do not claim that these critics replace human verification; rather, they
provide a practical baseline audit that can be tightened, inspected, or extended in future work.

\texttt{LLMGenQC} supports two complementary mechanisms.
\emph{Hard filtering} enforces minimum quality by rejecting candidates that fail any constraint (alignment,
safety, or consistency). This is useful when releasing synthetic corpora and when downstream systems are
sensitive to noisy metadata.
\emph{Soft control} reduces distribution shift among accepted texts by reweighting synthetic samples that
look less human-like in critic space, using density-ratio estimation.
Together, these mechanisms make the augmentation pipeline both conservative (via filtering) and adaptive
(via weighting), without requiring retraining of recommendation models.

All code for this section is released as a Colab notebook and can be run independently of the
large-scale experiments in Section~\ref{sec:benchmark_experiments}.

\subsection{Anchor construction and LLM generation}
\label{subsec:ragmovieaug_anchor}

For each movie $i$ we construct an \emph{anchor} $a_i$ by concatenating readily available metadata:
\begin{equation}
  a_i = \text{Title}_i \,\|\, \text{Year}_i \,\|\, \text{Genres}_i \,\|\, \text{Plot}_i \,\|\, \text{Tags}_i,
  \label{eq:anchor}
\end{equation}
where $\text{Tags}_i$ is the set of top-$K$ Tag Genome tags (if available) and $\text{Plot}_i$ is an optional
human-written synopsis. In practice we build two variants: a \emph{full} anchor $a_i^{\text{full}}$ that
includes all available fields, and a \emph{tag anchor} $a_i^{\text{tags}}$ that only contains tags and/or
genres.

Conditioned on $a_i$ we query an LLM via OpenRouter to obtain $K$ synthetic descriptions
\begin{equation}
  s_{ik} \sim q_\theta(\,\cdot \mid a_i\,), \quad k = 1,\dots,K,
  \label{eq:llm-samples}
\end{equation}
where $q_\theta$ denotes the LLM. The system prompt enforces \emph{anchor-constrained} behaviour:
the model is instructed to only use information implied by the anchor and to output the special token
\texttt{INSUFFICIENT} when the anchor does not support a faithful synopsis. We use two output styles
(paragraph synopses and aspect-style bullet lists) as complementary formats. In the toy study we sample a
small subset of MovieLens items and generate up to two candidates per item from different backbone models
(GPT--4o and Claude~3.5 in our implementation).

\subsection{Critic models and filtering}
\label{subsec:ragmovieaug_critics}

Each candidate description $s_{ik}$ is evaluated by a collection of pretrained \emph{critic models}
that approximate complementary quality dimensions (semantic faithfulness, fluency, safety, and consistency):

\begin{itemize}
  \item \textbf{SBERT relevance}
  \(
    c^\text{rel}_{ik}
    = \cos\big(f_\text{SBERT}(a_i^{\text{full}}), f_\text{SBERT}(s_{ik})\big)
  \)
  measures semantic alignment with the full anchor.
  \item \textbf{Tag agreement}
  \(
    c^\text{tag}_{ik}
    = \cos\big(f_\text{SBERT}(a_i^{\text{tags}}), f_\text{SBERT}(s_{ik})\big)
  \)
  emphasizes agreement with Tag Genome tags and genres (useful when plots are missing).
  \item \textbf{Fluency} is approximated by GPT--2 perplexity $c^\text{ppl}_{ik}$.
  \item \textbf{Toxicity} $c^\text{tox}_{ik} \in [0,1]$ is estimated with Detoxify.
  \item \textbf{Length} $c^\text{len}_{ik}$ is the word count of $s_{ik}$.
  \item \textbf{Consistency} is captured by an NLI-based \emph{contradiction rate}
  $c^\text{nli}_{ik}$, defined as the fraction of sentences in $s_{ik}$ that a BART-MNLI classifier labels as
  \textsc{Contradiction} with respect to the anchor $a_i^{\text{full}}$.
\end{itemize}

We collect these signals into a critic feature vector
\begin{equation}
  \phi_{ik}
  =
  \big[c^\text{rel}_{ik},
       c^\text{tag}_{ik},
       c^\text{ppl}_{ik},
       c^\text{tox}_{ik},
       c^\text{len}_{ik},
       1 - c^\text{nli}_{ik}
  \big]^\top.
  \label{eq:critic-vector}
\end{equation}

\paragraph{Hard filtering (accept/reject).}
A candidate $s_{ik}$ is accepted if all critic scores satisfy threshold constraints $\tau$:
\begin{equation}
  \mathsf{accept}(s_{ik}) =
  \mathbb{I}\Big[
    c^\text{rel}_{ik} \ge \tau_\text{rel},\;
    c^\text{tag}_{ik} \ge \tau_\text{tag},\;
    c^\text{tox}_{ik} \le \tau_\text{tox},\;
    \tau^\text{min}_\text{len} \le c^\text{len}_{ik} \le \tau^\text{max}_\text{len},\;
    c^\text{ppl}_{ik} \le \tau_\text{ppl},\;
    c^\text{nli}_{ik} \le \tau_\text{nli}
  \Big].
  \label{eq:critic-thresholds}
\end{equation}
This conservative rule is intended for quality assurance in released resources: it prunes off-topic or unsafe
generations before they can influence downstream retrieval or recommendation.

In the Colab notebook, we compute and store all critic scores in a CSV file, together with a pass/fail
flag for each candidate. This enables transparent auditing: for example, histograms of SBERT similarity and
contradiction scores for accepted versus rejected texts, and manual inspection of borderline cases.

\subsection{Toy quality analysis}
\label{subsec:ragmovieaug_toy}

The goal of this section is not to optimize recommender accuracy, but to provide a transparent
\emph{quality audit} of LLM-generated synopses that can be executed independently of the full benchmark.
After running the augmentation pipeline on a random subset of MovieLens items, the notebook:

\begin{enumerate}
  \item reports the overall \emph{acceptance rate} (fraction of LLM outputs that pass
        Eq.~\eqref{eq:critic-thresholds});
  \item compares the distributions of $c^\text{rel}$, $c^\text{tag}$, $c^\text{ppl}$,
        $c^\text{tox}$, $c^\text{nli}$ and $c^\text{len}$ between accepted and rejected candidates;
  \item groups movies by popularity (\emph{cold}, \emph{warm}, \emph{hot}) based on interaction counts,
        and reports mean acceptance rates per bucket;
  \item runs a small sensitivity analysis that varies key thresholds (e.g., $\tau_\text{rel}$ and
        $\tau_\text{nli}$) and records the resulting pass rate.
\end{enumerate}

These descriptive statistics verify that accepted synopses remain semantically close to the anchors
and exhibit low measured contradiction and toxicity, while also testing whether quality control behaves
differently in the long tail. All critic scores and thresholds are logged to disk so that other researchers
can adopt stricter or more permissive configurations, or inspect specific items.

\subsection{Density-ratio estimation}
\label{subsec:ragmovieaug_dr}

\paragraph{Soft control via reweighting.}
Beyond hard filtering, we further align the distribution of accepted synthetic texts with human-written
plots by estimating a density ratio in critic space. We collect pairs $(\phi, y)$ where $y = 1$ for human
texts (plots or anchors) and $y = 0$ for accepted synthetic texts, and fit a logistic regression model
\begin{equation}
  P_\theta(y=1 \mid \phi) = \sigma(w^\top \phi + b),
  \label{eq:logistic}
\end{equation}
with $\sigma(\cdot)$ the logistic function. Under standard assumptions, this yields an estimate of the
density ratio between human and synthetic distributions:
\begin{equation}
  r(\phi)
  \approx
  \frac{p_\text{human}(\phi)}{p_\text{synth}(\phi)}
  \propto
  \frac{P_\theta(y=1 \mid \phi)}{P_\theta(y=0 \mid \phi)}
  \cdot
  \frac{\pi_s}{\pi_h},
  \label{eq:density-ratio}
\end{equation}
where $\pi_s$ and $\pi_h$ are prior mixture proportions. In practice we use $r(\phi)$ as an importance
weight for each accepted synthetic description, clipped to a bounded interval for numerical stability.
This assigns higher influence to synthetic texts whose critic signatures resemble human plots.

\subsection{Adaptive mixing and cold-start emphasis}
\label{subsec:ragmovieaug_mixing}

For each movie $i$ we construct a mixture of human and synthetic texts.
Let $\mathcal{H}_i$ denote the set of human texts (at most one plot) and $\mathcal{S}_i$ the set of
accepted synthetic descriptions. We compute an \emph{anchor agreement} score
\begin{equation}
  \alpha_i
  =
  \frac{1}{2}\cos\big(f_\text{SBERT}(a_i^{\text{full}}), f_\text{SBERT}(h_i)\big)
  + \frac{1}{2}\big(1 - \bar{c}^\text{nli}_i\big),
  \label{eq:anchor-agree}
\end{equation}
where $h_i \in \mathcal{H}_i$ (if available) and $\bar{c}^\text{nli}_i$ is the average contradiction rate
among synthetic candidates for item $i$. We also compute the number of interactions $n^\text{inter}_i$ for
movie $i$ in MovieLens.

An adaptive mixing weight $\lambda_i \in [0.15, 0.85]$ is then computed as a function of
$n^\text{inter}_i$ and $\alpha_i$, assigning higher $\lambda_i$ to cold-start items (small $n^\text{inter}_i$)
and weaker anchor agreement. The final textual mixture for movie $i$ is
\begin{equation}
  \mathcal{T}_i = \{(t, w_t)\}_{t \in \mathcal{H}_i \cup \mathcal{S}_i},
  \label{eq:text-mixture}
\end{equation}
with weights
\begin{align}
  \sum_{t \in \mathcal{H}_i} w_t &= 1 - \lambda_i, \label{eq:mix-human} \\
  \sum_{t \in \mathcal{S}_i} w_t
  &= \lambda_i \cdot
     \frac{r(\phi_t)}{\sum_{s \in \mathcal{S}_i} r(\phi_s)},
  \label{eq:mix-synth}
\end{align}
where $r(\phi_t)$ is the density-ratio estimate from Eq.~\eqref{eq:density-ratio}. In the released
Colab notebook we store $\mathcal{T}_i$ together with all critic scores, enabling downstream work to plug these mixtures into embedding models, or to further adjust cold-start emphasis.

\begin{figure}
    \centering
\begin{tikzpicture}[
    node distance=7mm and 10mm, 
    auto,
    >=Stealth,
    block/.style={
        rectangle,
        draw,
        rounded corners,
        text width=3.2cm,
        align=center,
        minimum height=7mm,
        font=\small
    },
    arrow/.style={->, thick}
]
\node[block] (ml) {MovieLens data};
\node[block, below=of ml] (anchor) {Anchor\\$a_i$};
\node[block, below=of anchor] (llm) {LLM gen.\\$s_{ik} \sim q_\theta(\cdot \mid a_i)$};
\node[block, below=of llm] (critics) {Critics\\SBERT, GPT-2,\\Detoxify, MNLI};

\node[block, below left=of critics] (density) {Density ratio\\$r(\phi)$};
\node[block, below right=of critics] (mix) {Adaptive mix\\$\lambda_i, \mathcal{T}_i$};

\node[block, below=of mix] (recsys) {Recommender\\(item encoder, MF, \dots)};

\draw[arrow] (ml) -- (anchor);
\draw[arrow] (anchor) -- (llm);
\draw[arrow] (llm) -- (critics);
\draw[arrow] (critics) -- (density);
\draw[arrow] (critics) -- (mix);
\draw[arrow] (density) -- (mix);
\draw[arrow] (mix) -- (recsys);
\end{tikzpicture}
\end{figure}

\subsection{Discussion}

\texttt{LLMGenQC} is deliberately framed as a toy study: it operates on a subset of MovieLens items,
does not require retraining any recommendation model, and can be executed quickly on a standard Colab
instance. Nevertheless, it directly addresses common concerns about LLM-based augmentation.
Anchor-constrained prompting and SBERT-based alignment checks reduce semantic drift; NLI-based contradiction
rates provide an automated proxy for consistency with the anchor; and Detoxify provides a safety screen.
While this work does not include manual human annotation, all synthetic texts are released together with
critic scores and pass/fail decisions, enabling transparent inspection and threshold tightening when needed.
We return to empirical findings from a 500-movie instantiation in Section~\ref{sec:ragmovieaug_results}.

\section{Proposed Resource and System Implementation}
\label{sec:proposed_resource}

We present a novel resource for multimodal recommendation, integrating (i) LLM-generated textual descriptions and (ii) visual trailer embeddings into a retrieval-augmented pipeline. We detail our strategy for augmenting sparse movie metadata, summarize visual embedding extraction and fusion, and explain how these multimodal features support retrieval-augmented generation (RAG) and optional collaborative filtering (CF).

\subsubsection*{Overview and Motivation.}
\label{sec:resource_overview}

The proposed resource addresses sparse textual metadata by using LLMs to generate short movie descriptions, which are then embedded using modern text encoders (e.g., GPT, Sentence Transformers). Additionally, we use movie trailers to generate visual embeddings. Combining these textual and visual signals mitigates cold-start issues and captures cinematic nuances that text-only methods overlook. The resource content and main highlights include:
\begin{itemize}
    \item \textbf{Enriched Textual Descriptions:} LLMs expand sparse metadata into short paragraph-level synopses for each movie.
    \item \textbf{Visual Trailer Embeddings:} We adopt existing trailer-based embeddings (e.g., from MMTF-14K~\cite{mmtf14k}) or optionally generate them via CNN/Transformer backbones.
    \item \textbf{Fusion Methods:} We provide code for combining text and visual embeddings with PCA or CCA, enabling a single multimodal representation for retrieval or CF.
    \item \textbf{RAG and CF Integration:} We offer scripts to (1) perform embedding-based retrieval + LLM re-ranking (RAG), and (2) (optional) incorporate side embeddings into CF methods.\footnote{\url{https://anonymous.4open.science/r/RAG-VisualRec-2866/}} We also include routines for standard accuracy metrics, beyond-accuracy measures (novelty, tail coverage), and ablation analyses (temporal vs.\ average embeddings).
\end{itemize}
Overall, our resource unifies textual augmentation with visual data in a single, flexible environment, simplifying the design of multimodal recommenders for the movie domain.

\subsubsection*{Text Processing and Data Augmentation.}
\label{sec:data_augmentation}

To enrich the metadata of each film, we use a large language model with minimal inputs (\textit{title, genre, tags}) and a concise plot-driven description. For instance:
\begin{tcolorbox}[colback=gray!10,colframe=gray!50,boxrule=0.4pt]
\footnotesize
\textbf{Prompt Example:} 
\emph{``Given the title \{title\} and genres \{genre\}, write a short paragraph summarizing the film’s plot, themes, and style.''}
\end{tcolorbox}

The LLMs' output is saved as an \texttt{augmented\_description}, serving as a crucial textual feature when official synopses are missing. Table~\ref{table:enrichment_example_nixon} shows the enriched description for the movie \emph{Nixon (1995)}. We then produce embeddings from the augmented text via chosen encoders (e.g., OpenAI or Sentence Transformers), yielding a vector \(\mathbf{x}_i^{(\mathrm{txt})} \in \mathbb{R}^{d_\mathrm{txt}}\) for each item \(i\).

\begin{table}[!ht]
\centering
\small
\caption{Sample data augmentation for \textit{Nixon (1995)}. Minimal metadata is automatically expanded into a cohesive description highlighting historical context, thematic depth, and key figures.}
\label{table:enrichment_example_nixon}
\begin{tabular}{@{}l@{\quad}l@{\quad}p{7.3cm}@{}}
\toprule
\textbf{Aspect} & \textbf{Before} & \textbf{After LLM Augmentation} \\ 
\midrule
Title & Nixon (1995) & \textit{unchanged} \\
Genres & Drama \,|\, Biography & \textit{unchanged} \\
Description & \textit{Not provided} & 
\textit{``Nixon (1995) explores the troubled psyche and political career of America's 37th president, delving into both his strategic brilliance and the moral compromises that shaped his legacy. Directed by Oliver Stone, the film highlights Nixon's ...''} \\
\bottomrule
\end{tabular}
\end{table}

\subsubsection*{Visual Embeddings and Fusion Methods}
\label{sec:visual_fusion}
We incorporate trailer-level embeddings from publicly available datasets (e.g., MMTF-14K). Each trailer is sampled at fixed intervals, passed through a CNN or Vision Transformer, and aggregated (e.g., mean pooling) to produce $\mathbf{x}_i^{(\mathrm{vis})} \in \mathbb{R}^{d_\mathrm{vis}}$. We provide scripts for custom extraction if desired.

\paragraph{Multimodal Fusion.}
Since textual and visual signals can be complementary, we apply two main fusion strategies:
\begin{itemize}
    \item \textbf{Concatenation + PCA:} Stack $\mathbf{x}_i^{(\mathrm{txt})}$ and $\mathbf{x}_i^{(\mathrm{vis})}$, then project down to a fused vector $\mathbf{z}_i$ via PCA.
    \item \textbf{CCA Alignment:} Learn a shared subspace maximizing cross-modal correlation, yielding aligned vectors for text and visuals. These aligned components are concatenated as the fused embedding.
\end{itemize}
Either fused or unimodal embeddings are then fed into the recommender.
\subsubsection*{Integration into RAG and CF}
We adopt the two-step retrieval and LLM-based re-ranking pipeline described in Section~\ref{sec:formal}: (i) retrieve candidates via $k$-NN in user-aggregated embedding space, and (ii) re-rank these candidates using LLM prompts informed by user profiles and candidate metadata. In our experiment, we fix \( k = 10 \) and vary the retrieval scope \( N \in [20, 30, 50, 100, 150] \) to analyze its impact (cf. {\hyperref[sec:rq4]{RQ4}}).

This approach harnesses the immediate semantic alignment of embeddings for retrieval, while the final list is refined by the LLM, enhancing personalization and reducing hallucinations (since items are anchored to known vectors).

\begin{tcolorbox}[colback=lightgray!15, colframe=gray!50, boxrule=0.5pt, sharp corners]
\textbf{Note.} This paper introduces multimodal data tailored for evaluating RAG-based methods in RSs. While experiments integrating side information embeddings into CF models (e.g., visual embeddings via VBPR or ConvMF, textual embeddings via CTR) are deferred to future work, example implementations using the \texttt{Cornac} \cite{cornac} library are provided in the GitHub repository.
\end{tcolorbox}

\subsubsection*{Evaluation Metrics and Hyper-Parameter Settings}

\paragraph{Accuracy metrics at two cut-offs.}
We report ranking accuracy at two stages.
\begin{itemize}
    \item \textbf{Retrieval stage:} we evaluate the top-$K$ candidate set returned by kNN retrieval and report
Recall@$K$ and nDCG@$K$.
\item \textbf{Recommendation stage:} we evaluate the final top-$K$ list output by the generation module
(LLM-based selection/re-ranking) and report Recall@$K$ and nDCG@$K$.
Unless stated otherwise, we fix $K=10$ and vary $N \in \{20,30,50,100,150\}$.
\end{itemize}

Accuracy results are reported as Recall@k and nDCG@k, computed in
the standard way over the per-user top-$k$ list $\mathcal{R}_u$. Let $\mathcal{T}_u$ denote
the set of relevant test items for user $u$. Then
\[
\mathrm{Recall@k}(u)
  = \frac{|\mathcal{R}_u \cap \mathcal{T}_u|}{|\mathcal{T}_u|}, \qquad
\mathrm{nDCG@k}(u)
  = \frac{1}{\mathrm{IDCG@k}(u)}
    \sum_{i \in \mathcal{R}_u}
      \frac{\mathbb{I}[i \in \mathcal{T}_u]}{\log_2(1 + \mathrm{rank}_u(i))}.
\]
We report the average of these per-user scores across the test users.

\paragraph{Beyond-Accuracy metrics.}

Beyond standard ranking metrics such as Recall and nDCG, we also measure diversity and long-tail aspects:

\noindent
\textbf{Item Coverage} is defined inline as 
\(\mathrm{Coverage} = \tfrac{\bigl|\bigcup_{u\in\mathcal{U}} \mathcal{R}(u)\bigr|}{|\mathcal{I}|}\),
where \(\mathcal{R}(u)\) is the top-\(k\) recommendation list for user \(u\) and \(\mathcal{I}\) is the set of all items.\vspace{0.9mm}

\noindent
\textbf{Novelty} measures the average unexpectedness of recommended items:
\(\mathrm{Novelty} = \tfrac{1}{|\mathcal{R}|} \sum_{i\in \mathcal{R}} -\log_2\bigl(p(i)\bigr)\),
where \(p(i)\) is the relative popularity of item \(i\). Higher values indicate recommendations of less popular (thus more novel) items. \vspace{0.9mm}

\noindent
\textbf{LongTail Fraction} is given by $\mathrm{LongTailFrac} = \frac{1}{|\mathcal{R}|} \sum_{i\in \mathcal{R}} \mathbf{1}\bigl[\text{($i$ is tail)}\bigr]$ where an item is ``tail'' if it falls below a certain popularity threshold ({in this work, \( \tau_{\text{tail}}=2 \)}).

\noindent
\textbf{Hyper-Parameter Settings.} The following describes our choice of hyperparameters:
\begin{itemize}
    \item \textbf{LLM Generation}: Temperature \(=0.7\), max tokens \(=200\).
    \item \textbf{\(k\)-NN retrieval}: \(N=50\) or \(N=100\) neighbors; final top-\(k=10\) for recommendations.
    \item \textbf{PCA or CCA dims}: \(d_z=64\) or \(128\), chosen via grid search.
    \item \textbf{CF Training}: 10 latent dimensions, learning rate \(=0.01\), up to 5--10 epochs, with a 70:30 train-test split.
\end{itemize}
Overall, these choices provide baseline settings for reproducibility; finer hyperparameter tuning is possible as needed.

\textbf{Dataset}. We benchmark our approach using the MovieLens (Latest) dataset \cite{movielens}, filtering out users {between 20 to 100} interactions. We perform a chronological train–test split per user, allocating the earliest 70\% of interactions for training and the latest 30\% for testing. The dataset contains $100,836$ interactions, \(|U|=610\), \(|I|=9,724\), average interactions per user $165.30$, per item $10.37$, and sparsity of $0.017$. For efficiency, the evaluation uses $120$ randomly selected users. Items are categorized into head (popular), mid-tail, or long-tail based on rating frequency.

\begin{table*}[t!]
    \centering
    \caption{Retrieval-stage results measured as Recall@$N$ and nDCG@$N$ on the top-$N$ retrieved candidates
(here $N=50$ unless stated). “temp” and “avg” denote temporal vs.\ average user embeddings. The best values are highlighted in green, while the second-best are in yellow. $\Delta_{NDCG}$ indicates the relative improvement in NDCG (temp) over the baseline.}

    \label{tab:retrieval_stage_final}
    \resizebox{\textwidth}{!}{
        \begin{tabular}{lcc|cc|cccc|c}
        \toprule
            \textbf{Name} & \multicolumn{2}{c|}{\textbf{Uni/Multimodal}} & \multicolumn{2}{c|}{\textbf{Fusion}} & \multicolumn{4}{c|}{\textbf{Metrics}} & \multirow{2}{*}{\textbf{$\Delta_{ndcg}$}} \\
            \cmidrule(lr){2-3}\cmidrule(lr){4-5}\cmidrule(lr){6-9}
             & Visual & Text & PCA & CCA & Recall (temp) & NDCG (temp) & Recall (avg) & NDCG (avg) & \\
        \midrule
            Visual Unimodal     & \checkmark &         &      &      & 0.036688          & 0.037534          & 0.026457          & 0.025499   & $\pm0.0\%$       \\
        \midrule
            ST Unimodal         &      & \checkmark &      &      & 0.160541          & 0.159737          & \secondbest{0.116791} & 0.089738    & $+325.6\%$      \\
            OpenAI Unimodal     &      & \checkmark &      &      & \secondbest{0.176216} & \secondbest{0.181286} & 0.116251          & \secondbest{0.101413} & $+383.0\%$ \\
            Llama3.0 Unimodal   &      & \checkmark &      &      & 0.093495          & 0.096862          & 0.070260          & 0.055245   & $+158.1\%$       \\
        \midrule
            Visual+Textual (ST)
                                & \checkmark & \checkmark & \checkmark &      & 0.104287          & 0.122746          & 0.092985          & 0.096431    & $+227.6\%$      \\
            Visual+Textual (OpenAI)
                                & \checkmark & \checkmark & \checkmark &      & 0.104882          & 0.123015          & 0.093541          & 0.096444    & $+228.4\%$      \\
            Visual+Textual (Llama3.0) 
                                & \checkmark & \checkmark & \checkmark &      & 0.108265          & 0.139158          & 0.089097          & 0.099020    & $+271.3\%$      \\
            Visual+Textual (ST) 
                                & \checkmark & \checkmark &      & \checkmark & \best{0.205637}   & \best{0.252080}   & \best{0.180762}   & \best{0.174588} & $+571.6\%$  \\
            Visual+Textual (OpenAI) 
                                & \checkmark & \checkmark &      & \checkmark & 0.087140          & 0.080328          & 0.072103          & 0.055052    & $+114.0\%$      \\
            Visual+Textual (Llama3.0) 
                                & \checkmark & \checkmark &      & \checkmark & 0.119989          & 0.148211          & 0.099557          & 0.095724    & $+294.9\%$      \\
        \bottomrule
        \end{tabular}
    }
\end{table*}

\section{Benchmark and Experiments}
\label{sec:benchmark_experiments}

Table~\ref{tab:retrieval_stage_final} and Table~\ref{tab:recommendation_stage_final} summarize the key findings for both \textbf{Retrieval} and \textbf{Recommendation} stages, contrasting unimodal (\texttt{Visual Unimodal}, \texttt{ST Unimodal}, \texttt{OpenAI Unimodal}, \texttt{Llama3.0 Unimodal}) and multimodal (\texttt{Visual+Textual} with \texttt{PCA} or \texttt{CCA}). Although we observe broadly similar trends in retrieval and recommendation, our primary focus is on the \emph{Recommendation Stage} results (Table~\ref{tab:recommendation_stage_final}), as the final ranking is most relevant for practical systems. We nonetheless reference the \emph{Retrieval Stage} (Table~\ref{tab:retrieval_stage_final}) to confirm that both stages share consistent patterns. To organize our discussion, we define four experimental research questions (\texttt{RQ1}--\texttt{RQ4}). Note that the terms \textbf{Manual} and \textbf{LLM-based} pipelines specifically refer to the augmentation process (Sec. \ref{sec:formal}).

\begin{table*}[ht!]
    \centering
    \caption{Recommendation-stage results measured as Recall@K and nDCG@K on the final top-$K$ list ($K=10$) returned by the generation module.  We use the same color-based notation and columns as in Table~\ref{tab:retrieval_stage_final}.}
    \label{tab:recommendation_stage_final}
    \resizebox{\textwidth}{!}{%
    \begin{tabular}{lcc|cc|cccc|c}
    \toprule
        \multicolumn{9}{l}{\textbf{Recommendation Stage -- Manual Pipeline}} \\
    \midrule
        \textbf{Name} & \multicolumn{2}{c|}{\textbf{Uni/Multimodal}} & \multicolumn{2}{c|}{\textbf{Fusion}} & \multicolumn{4}{c|}{\textbf{Metrics}} & \multirow{2}{*}{\textbf{$\Delta_{ndcg}$}} \\
    \cmidrule(lr){2-3}\cmidrule(lr){4-5}\cmidrule(lr){6-9}
     & Visual & Text & PCA & CCA & Recall (temp) & NDCG (temp) & Recall (avg) & NDCG (avg) & \\
    \midrule
        Visual Unimodal     & \checkmark &         &      &      & 0.01611          & 0.045019          & 0.01198          & 0.030501    & $\pm0.0\%$      \\
    \midrule
        ST Unimodal         &      & \checkmark &      &      & \secondbest{0.084321}          & \secondbest{0.167488}          & 0.050541          & 0.079941    & $+272.0\%$      \\
        OpenAI Unimodal     &      & \checkmark &      &      & 0.075686 & 0.166709 & \secondbest{0.059188} & \secondbest{0.104077} & $+270.3\%$ \\
        Llama3.0 Unimodal   &      & \checkmark &      &      & 0.03982          & 0.091084          & 0.034442          & 0.062854  & $+102.3\%$        \\
    \midrule
        Visual+Textual (ST) 
                            & \checkmark & \checkmark & \checkmark &      & 0.043374          & 0.12483          & 0.035712          & 0.080947     & $+177.3\%$     \\
        Visual+Textual (OpenAI) 
                            & \checkmark & \checkmark & \checkmark &      & 0.037374          & 0.115342          & 0.036908          & 0.078362   & $+156.2\%$      \\
        Visual+Textual (Llama3.0) 
                            & \checkmark & \checkmark & \checkmark &      & 0.037333          & 0.114743          & 0.032583          & 0.075561    & $+154.9\%$      \\
        Visual+Textual (ST) 
                            & \checkmark & \checkmark &      & \checkmark & \best{0.107084}   & \best{0.268102}   & \best{0.095418}   & \best{0.174363}  & $+495.5\%$ \\
        Visual+Textual (OpenAI) 
                            & \checkmark & \checkmark &      & \checkmark & 0.028301          & 0.078006          & 0.025079          & 0.059463    & $+73.3\%$      \\
        Visual+Textual (Llama3.0) 
                            & \checkmark & \checkmark &      & \checkmark & 0.047614          & 0.145944          & 0.041977          & 0.09262     & $+224.2\%$    \\
    \midrule
    \multicolumn{9}{l}{\textbf{Recommendation Stage -- LLM-based Pipeline}} \\
    \midrule
        \textbf{Name} & \multicolumn{2}{c|}{\textbf{Uni/Multimodal}} & \multicolumn{2}{c|}{\textbf{Fusion Method}} & \multicolumn{4}{c|}{\textbf{Metrics}} & \multirow{2}{*}{\textbf{$\Delta_{ndcg}$}} \\
        \cmidrule(lr){2-3}\cmidrule(lr){4-5}\cmidrule(lr){6-9}
         & Visual & Text & PCA & CCA & Recall (temp) & NDCG (temp) & Recall (avg) & NDCG (avg) \\
    \midrule
        Visual Unimodal     & \checkmark &         &      &      & 0.01734          & 0.046774          & 0.00847          & 0.027337    & $\pm0.0\%$      \\
    \midrule
        ST Unimodal         &      & \checkmark &      &      & 0.0773          & 0.169313          & 0.051284          & 0.086847  & $+262.0\%$        \\
        OpenAI Unimodal     &      & \checkmark &      &      & \secondbest{0.089671} & \secondbest{0.179811} & \secondbest{0.057714} & \secondbest{0.101169} & $+284.4\%$ \\
        Llama3.0 Unimodal   &      & \checkmark &      &      & 0.033492          & 0.083699          & 0.032395          & 0.066655     & $+78.9\%$     \\
    \midrule
        Visual+Textual (ST) 
                            & \checkmark & \checkmark & \checkmark &      & 0.03513          & 0.120745          & 0.030523          & 0.06666    & $+158.1\%$      \\
        Visual+Textual (OpenAI) 
                            & \checkmark & \checkmark & \checkmark &      & 0.036987          & 0.101464          & 0.034161          & 0.072735   & $+116.9\%$       \\
        Visual+Textual (Llama3.0) 
                            & \checkmark & \checkmark & \checkmark &      & 0.038048          & 0.118645          & 0.028749          & 0.078329     & $+153.7\%$     \\
        Visual+Textual (ST) 
                            & \checkmark & \checkmark &      & \checkmark & \best{0.097654}   & \best{0.246567}   & \best{0.089355}   & \best{0.159758} & $+427.1\%$  \\
        Visual+Textual (OpenAI) 
                            & \checkmark & \checkmark &      & \checkmark & 0.04025          & 0.084531          & 0.02919          & 0.063593    & $+80.7\%$      \\
        Visual+Textual (Llama3.0) 
                            & \checkmark & \checkmark &      & \checkmark & 0.048338          & 0.141869          & 0.034597          & 0.079539    & $+203.3\%$      \\
    \bottomrule
    \end{tabular}
    }
\end{table*}

\begin{description}[leftmargin=0.7cm,labelindent=0cm]
    \item[\texttt{RQ1}] \textit{Does incorporating(i.e., fusing) visual features improve accuracy metrics across different textual LLM backbones, and why is this crucial?}
    \item[\texttt{RQ2}] \textit{Which fusion method (\texttt{PCA} or \texttt{CCA}) is more effective for combining textual and visual embeddings, and why does this matter?}
    \item[\texttt{RQ3}] \textit{How do multimodal pipelines affect beyond-accuracy metrics (Coverage, Novelty, TailFrac), and why are these additional metrics important for user satisfaction?}
    \item[\texttt{RQ4}] \textit{What differences emerge between the \texttt{Manual} vs.\ \texttt{LLM-based} pipelines in final recommendation quality, and what is the impact of retrieval stage scope $N$?}
    \item[\texttt{RQ5}] How reliable are LLM-generated item synopses under critic-based auditing (\texttt{LLMGenQC}), and how does the acceptance rate vary across item popularity buckets (cold/warm/hot)?
    \item[\texttt{RQ6}] How do key pipeline design choices (retrieval depth \(N\), fusion settings, and user-embedding construction) affect accuracy and beyond-accuracy metrics?
\end{description}

Please note that in the table, we also report \(\Delta\), which measures the improvement over a similar visual baseline. However, in the text, we do not always explicitly reference this value. 

\subsection{\texttt{RQ1}: Impact of Visual Information Across LLM Backbones}
\label{sec:rq1}
This question is important because each textual LLM encoder (e.g., \texttt{ST}, \texttt{OpenAI}, \texttt{Llama3.0}) may capture different semantic nuances, and we want to know whether adding visual signals consistently improves performance across all backbones or primarily benefits certain encoders. Understanding this helps us decide when and how to invest in visual data extraction.

\paragraph{Findings.}
From Table~\ref{tab:recommendation_stage_final} (Manual Pipeline), \texttt{ST Unimodal} achieves about $0.0843$ in Recall(\texttt{temp}) and $0.1675$ in NDCG(\texttt{temp}), whereas \texttt{OpenAI Unimodal} follows closely (Recall(\texttt{temp}) $=0.0757$, NDCG(\texttt{temp}) $=0.1667$). \texttt{Llama3.0 Unimodal} lags behind with Recall(\texttt{temp}) $\approx 0.0398$. Purely visual embeddings alone (\texttt{Visual Unimodal}) are substantially lower, with recall $\approx 0.0161$.

When we fuse visual features, performance increases dramatically, particularly for \texttt{ST} and \texttt{OpenAI}. For example, \texttt{Visual+Textual (ST) CCA} raises recall(\texttt{temp}) to $\approx 0.1071$ (a jump of more than $+27\%$ from \texttt{ST Unimodal}) and NDCG(\texttt{temp}) from $0.1675$ to $0.2681$ (about $+60\%$). Similar improvements are seen at the retrieval stage (Table~\ref{tab:retrieval_stage_final}), confirming that visual data consistently helps across both coarse retrieval and final recommendation. These results suggest that \textbf{incorporating visual embeddings can be crucial for capturing cinematic or aesthetic factors} that purely text-based approaches may miss, and this benefit is especially pronounced for \texttt{ST} and \texttt{OpenAI} backbones.

\subsection{\texttt{RQ2}: Comparing \texttt{PCA} vs.\ \texttt{CCA} Fusion}
\label{sec:rq2}
This question matters because \texttt{PCA} and \texttt{CCA} represent two distinct philosophies of dimensionality reduction: \texttt{PCA} aligns with global variance in a concatenated space, whereas \texttt{CCA} explicitly models correlations between textual and visual embeddings. Choosing the right method can critically affect the synergy between text and visuals.

\paragraph{Findings.}
From Table~\ref{tab:recommendation_stage_final}, \texttt{PCA}-based fusion yields moderate gains over unimodal systems but is consistently outperformed by \texttt{CCA}. Accordingly, and in the Manual Pipeline, \texttt{Visual+Textual (ST) PCA} hits Recall(\texttt{temp}) $\approx 0.0434$ and NDCG(\texttt{temp}) $\approx 0.1248$, whereas \texttt{Visual+Textual (ST) CCA} surpasses $0.1071$ in recall and $0.2681$ in NDCG (more than double the \texttt{PCA} values). The retrieval-stage results (Table~\ref{tab:retrieval_stage_final}) echo this pattern, with \texttt{ST CCA} reaching up to $0.2056$ in recall(\texttt{temp}). Thus, \textbf{CCA-based} approaches appear far more adept at capturing cross-modal alignment, thereby boosting both recall and ranking quality. For practitioners, this implies that \emph{simply concatenating text and visuals into a single \texttt{PCA} space is suboptimal}; learning a correlation-maximizing transformation (\texttt{CCA}) consistently leads to better synergy.

\begin{table*}[ht!]
    \centering
    \scriptsize
    \caption{Recommendation Stage Results (Beyond Accuracy Metrics). The table is split into two parts:
    \textbf{Manual Pipeline} (top) and \textbf{LLM-based Pipeline} (bottom). 
    We use the same color-based notation as in Table~\ref{tab:retrieval_stage_final}.}
    \label{tab:recommendation_stage_ba}
    \resizebox{\textwidth}{!}{%
    \begin{tabular}{lcc|cc|ccc}
    \toprule
        \multicolumn{8}{l}{\textbf{Recommendation Stage -- Manual Pipeline}} \\
    \midrule
        \textbf{Name} & \multicolumn{2}{c|}{\textbf{Uni/Multimodal}} & \multicolumn{2}{c|}{\textbf{Fusion Method}} & \multicolumn{3}{c}{\textbf{Metrics}} \\
    \cmidrule(lr){2-3}\cmidrule(lr){4-5}\cmidrule(lr){6-8}
     & Visual & Text & PCA & CCA & Coverage & Novelty & TailFrac \\
    \midrule
        Visual Unimodal     & \checkmark &         &      &      & 0.165006   & \secondbest{11.303532}    & 0.048333  \\
    \midrule
        ST Unimodal         &      & \checkmark &      &      & 0.234122	          & 11.10709          & 0.1075          \\
        OpenAI Unimodal     &      & \checkmark &      &      & 0.216065 & 10.797482 & 0.043333 \\
        Llama3.0 Unimodal   &      & \checkmark &      &      & 0.200498  & 11.288529    & \best{0.101667}    \\
    \midrule
        Visual+Textual (ST) 
                            & \checkmark & \checkmark & \checkmark &      & 0.308219  & 11.162305  & 0.062189  \\
        Visual+Textual (OpenAI) 
                            & \checkmark & \checkmark & \checkmark &      & 0.305106     & 11.163761   & 0.068333  \\
        Visual+Textual (Llama3.0) 
                            & \checkmark & \checkmark & \checkmark &      & 0.273973  & 11.112711 & 0.065833  \\
        Visual+Textual (ST) 
                            & \checkmark & \checkmark &      & \checkmark & \secondbest{0.32254} & 10.931898  & 0.0625  \\
        Visual+Textual (OpenAI) 
                            & \checkmark & \checkmark &      & \checkmark & \best{0.345579}  & \best{11.320455} & \secondbest{0.090833}  \\
        Visual+Textual (Llama3.0) 
                            & \checkmark & \checkmark &      & \checkmark & 0.291407  & 11.128886  & 0.0825   \\
    \midrule
    \multicolumn{8}{l}{\textbf{Recommendation Stage -- LLM-based Pipeline}} \\
    \midrule
    \textbf{Name} & \multicolumn{2}{c|}{\textbf{Uni/Multimodal}} & \multicolumn{2}{c|}{\textbf{Fusion Method}} & \multicolumn{3}{c}{\textbf{Metrics}} \\
    \cmidrule(lr){2-3}\cmidrule(lr){4-5}\cmidrule(lr){6-8}
     & Visual & Text & PCA & CCA & Coverage & Novelty & TailFrac \\
    \midrule
        Visual Unimodal     & \checkmark &         &      &      & 0.1401 & \secondbest{11.29123}  & 0.0475  \\
    \midrule
        ST Unimodal         &      & \checkmark &      &      & 0.202366  & 11.152372  & \best{0.096667}  \\
        OpenAI Unimodal     &      & \checkmark &      &      & 0.167497 & 10.700434 & 0.023333 \\
        Llama3.0 Unimodal   &      & \checkmark &      &      & 0.156912 & 11.128583  & 0.079167  \\
    \midrule
        Visual+Textual (ST) 
                            & \checkmark & \checkmark & \checkmark &      & 0.281445  & 11.137257  & 0.06 \\
        Visual+Textual (OpenAI) 
                            & \checkmark & \checkmark & \checkmark &      & 0.272105 & 11.061737  & 0.050833   \\
        Visual+Textual (Llama3.0) 
                            & \checkmark & \checkmark & \checkmark &      & 0.256538 & 10.950571 & 0.045   \\
        Visual+Textual (ST) 
                            & \checkmark & \checkmark &      & \checkmark & \secondbest{0.314446}  & 10.815412 & 0.064167 \\
        Visual+Textual (OpenAI) 
                            & \checkmark & \checkmark &      & \checkmark & \best{0.328144} & \best{11.335649}  & \secondbest{0.085833}  \\
        Visual+Textual (Llama3.0) 
                            & \checkmark & \checkmark &      & \checkmark & 0.278954  & 11.047586  & 0.070774 \\
    \bottomrule
    \end{tabular}
    }
\end{table*}

\subsection{\texttt{RQ3}: Influence on Coverage, Novelty, and TailFrac}
\label{sec:rq3}
Beyond-accuracy metrics are increasingly crucial in recommender systems because users often value diversity, discover less popular items, and receive suggestions that go beyond the mainstream. Here, we examine how multimodal pipelines impact \emph{Coverage}, \emph{Novelty}, and \emph{TailFrac} (share of long-tail items).

\begin{figure}[!t]
    \centering
    \includegraphics[width=\linewidth]{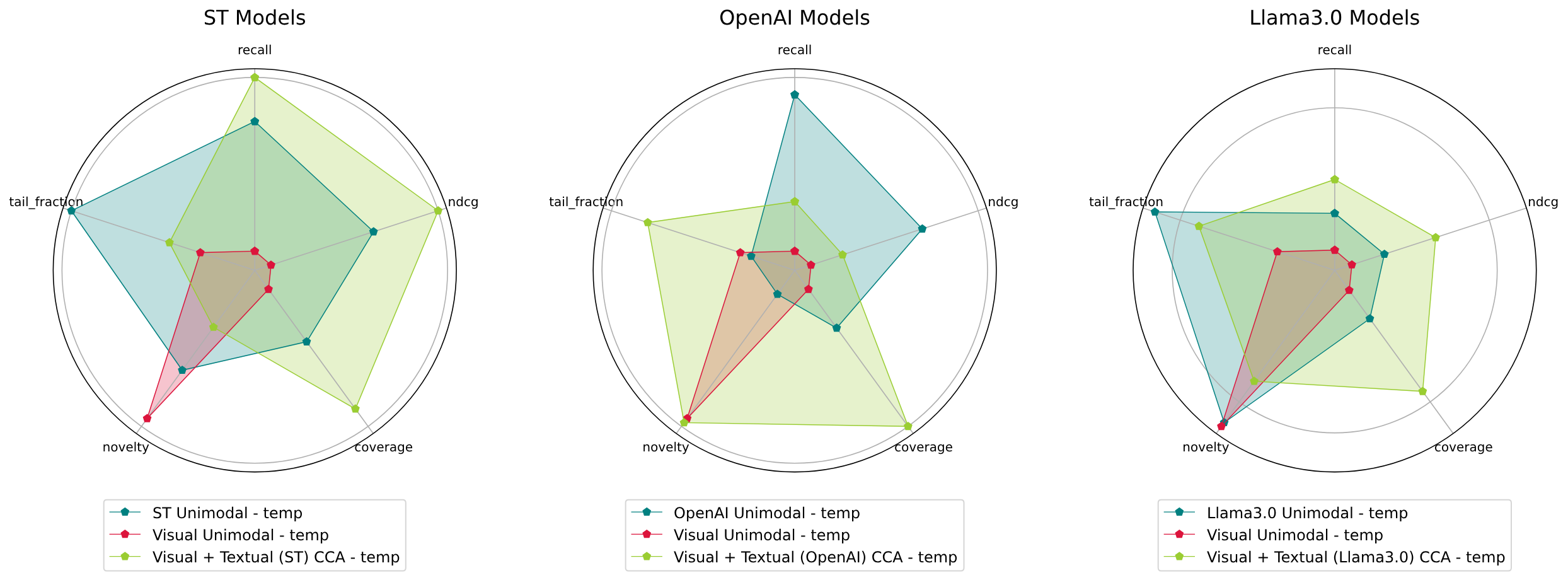}
\caption{Radar plot illustrating the performance of the recommendation stage across both accuracy and beyond-accuracy metrics for various LLMs.}    \label{fig:radar_charts}
\end{figure}

\paragraph{Findings.}
The attached beyond-accuracy table (Table~\ref{tab:recommendation_stage_ba}) and radar charts (Fig.~\ref{fig:radar_charts}) show that fusing visual and textual embeddings generally raises coverage by $0.10$--$0.15$ in absolute terms compared to unimodal text, thus allowing more items to appear in recommendations. Novelty metrics (where higher indicates recommending less familiar or popular items) also increase, especially under \texttt{CCA} fusion. For example, \texttt{OpenAI Unimodal} might have coverage near $0.216$, whereas \texttt{OpenAI+Visual (CCA)} can exceed $0.32$--$0.34$. However, the \texttt{TailFrac} measure can slightly decline in some cases, as the system recommends more mid-popular items (rather than extremely obscure ones). In Fig.~\ref{fig:radar_charts}, fused methods consistently expand coverage and novelty axes relative to unimodal baselines, but the exact impact on deep-tail exposure varies. In summary, \textbf{multimodal pipelines typically broaden the system’s recommendation space}, improving item discovery and novelty, with a slight trade-off in extreme tail coverage for certain encoders.

\subsection{\texttt{RQ4}: Differences Between \texttt{Manual} and \texttt{LLM-based} Pipelines}
\label{sec:rq4}
This final question is relevant because many real-world systems use an \texttt{LLM}-assisted re-ranking stage to refine the top-$k$ items after an initial retrieval or manual pipeline. We want to see if \texttt{LLM}-based re-ranking substantially reshuffles the best methods or simply provides incremental gains.

\paragraph{Findings.}
From the bottom half of Table~\ref{tab:recommendation_stage_final}, we note that \texttt{LLM-based} re-ranking preserves the overall hierarchy of approaches (\texttt{ST-CCA}, \texttt{OpenAI-CCA} \textgreater\ \texttt{ST-PCA}, etc.), but it can add $+5\%$ to $+10\%$ improvement in recall or NDCG for unimodal text methods. For instance, \texttt{OpenAI Unimodal} sees NDCG(\texttt{temp}) rise from $0.1667$ (\texttt{Manual}) to $0.1798$ (\texttt{LLM-based}), while \texttt{Visual+Textual (ST) PCA} also jumps from around $0.1248$ in NDCG(\texttt{temp}) to above $0.1207$ or $0.130$ in some runs. Overall, \emph{the best-fused methods remain best} under both pipelines, but the \texttt{LLM-based} approach refines the ranking further, especially for weaker unimodal baselines. This indicates that while second-stage re-ranking can sharpen the final list, \textbf{strong multimodal representations remain the core factor} for high recommendation quality.

\begin{figure}
    \centering
    \includegraphics[width=\linewidth]{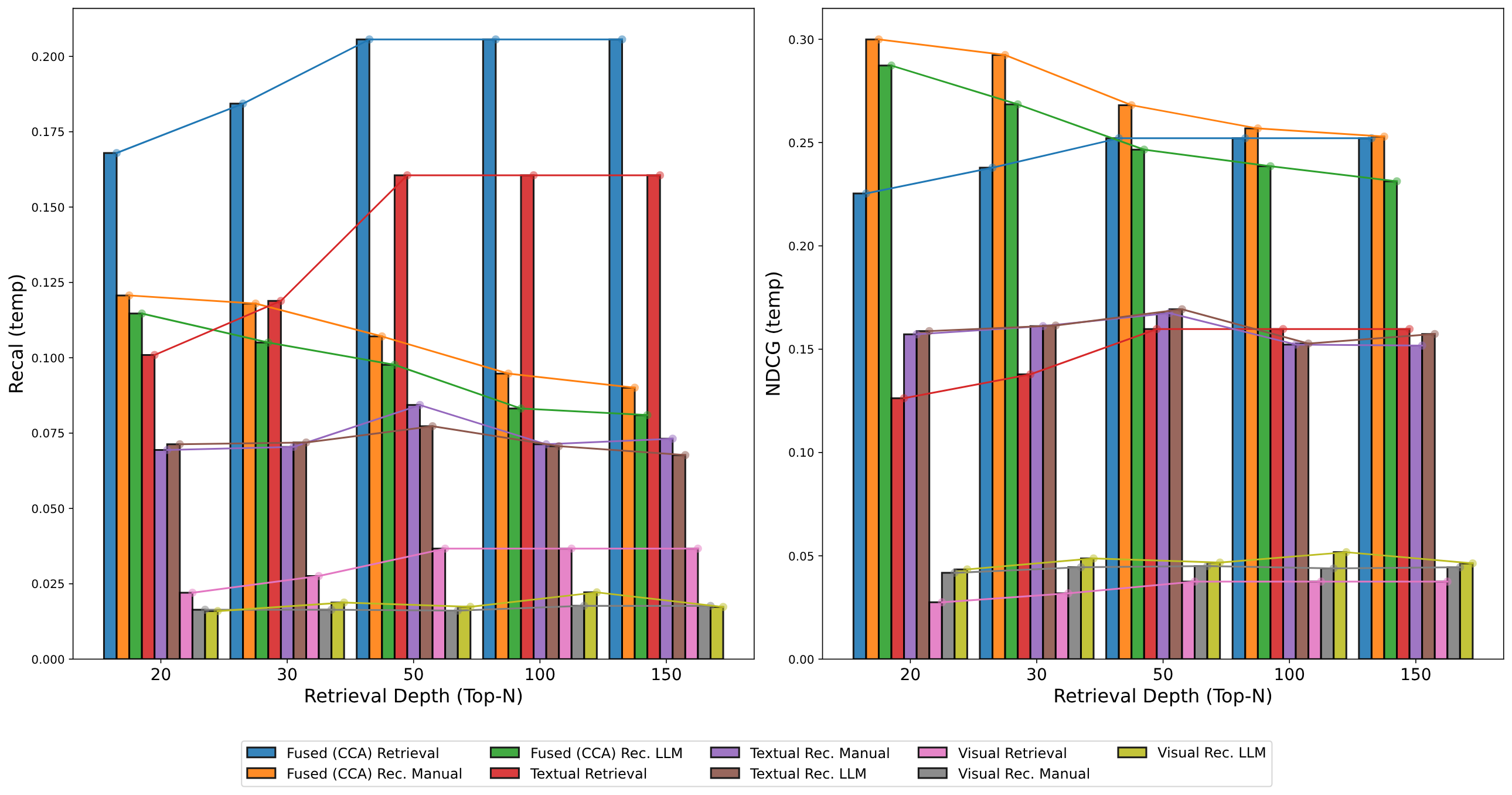}
    \caption{Ablation study across various retrieval depths (Top-N), evaluating temporal NDCG and recall values.}
    \label{fig:ablation}
\end{figure}

Regarding the impact of retrieval scope, in general, we find that an LLM-based re-ranking stage can further refine top-$k$ recommendations, especially for textual-scarce items. Meanwhile, varying the retrieval depth ($N=20$ to $N=150$) shows a well-known trade-off: broader retrieval may improve recall initially but eventually saturates or slightly harms nDCG due to noisier candidates. 
Figure~\ref{fig:ablation} (example ablation) confirms that certain fused pipelines (e.g., CCA-based) benefit from moderate $N$ increases, while purely text or purely visual retrieval sees diminishing returns. 

\subsection{RQ5: Quality of LLM-Generated Synopses (\texttt{LLMGenQC} Toy Study)}
\label{sec:ragmovieaug_results}

\paragraph{Goal and setup.}
We study whether LLM-generated item-side synopses are (i) semantically aligned with available metadata, (ii) internally consistent, and (iii) safe (non-toxic), and whether these properties differ across item popularity buckets. We operationalize this with \texttt{LLMGenQC}, an LLM-as-judge audit that produces per-synopsis critic scores and a binary decision label. Concretely, \emph{$\mathsf{pass}=1$ (accepted)} means a synopsis satisfies \emph{all} configured thresholds in Eq.~\eqref{eq:critic-thresholds}; \emph{$\mathsf{pass}=0$ (rejected)} means it fails at least one check (e.g., insufficient semantic alignment, excessive toxicity, or a detected contradiction).

For this toy study, we sampled 500 random movies, generated up to two candidate synopses per item
using two backbone models, and scored each candidate using the critic suite from
Section~\ref{sec:ragmovieaug}: SBERT similarity to the full anchor and to the tag-only anchor,
GPT--2 perplexity, Detoxify toxicity, NLI-based contradiction rate, and length.

\paragraph{How to read the critic scores.}
The most informative signals for anchor faithfulness are the SBERT-based similarities.
$c^{\text{rel}}$ measures semantic alignment with the \emph{full} anchor (title, year, genres, tags, optional plot),
while $c^{\text{tag}}$ focuses on agreement with genre and tag cues.
Higher values indicate that the synopsis stays closer to the intended movie semantics.
The remaining critics serve complementary roles: NLI contradiction and Detoxify flag inconsistency and unsafe language,
GPT--2 perplexity provides a fluency sanity check, and $c^{\text{len}}$ prevents trivial failures (extremely short or overly long outputs).

\paragraph{Accepted vs. rejected candidates (what $\mathsf{pass}$ means in practice).}
Roughly half of all candidates (49\%) passed all critic thresholds in Eq.~\eqref{eq:critic-thresholds}.
This acceptance rate should not be interpreted as low quality from the generator: the decision is intentionally conservative,
because it requires meeting \emph{multiple} constraints at the same time.
A single weak dimension is sufficient to mark a synopsis as $\mathsf{pass}=0$ so that off-topic or unreliable generations
are pruned before downstream use, yielding a high-precision and auditable synthetic corpus.

\begin{figure}[t]
  \centering
  \includegraphics[width=0.95\linewidth]{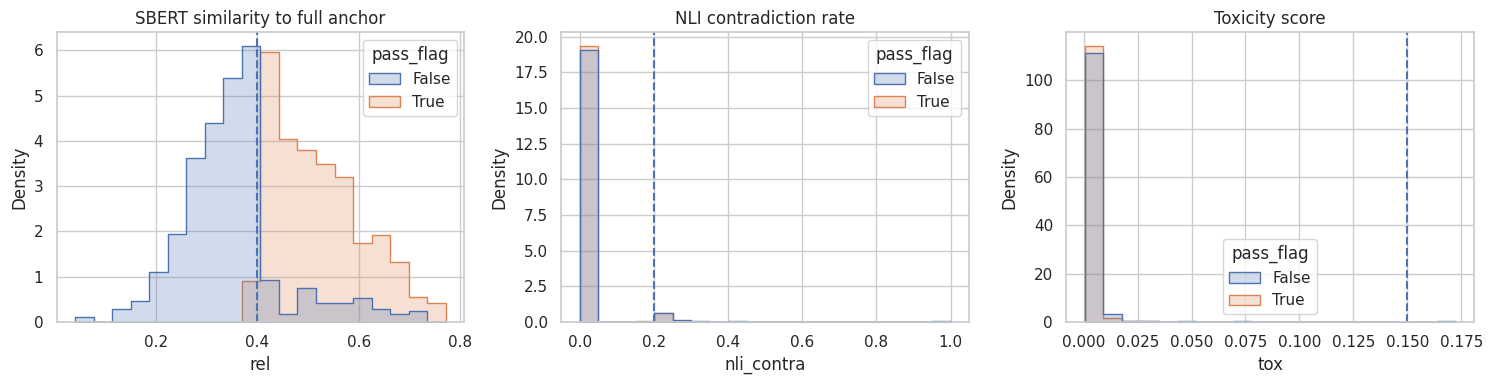}
  \caption{\textbf{LLMGenQC critic-score distributions in the 500-movie toy study, split by the final accept/reject decision.}
Histograms show (left) SBERT semantic similarity to the full anchor $a_i$, (middle) NLI-based contradiction rate, and (right) Detoxify toxicity score, each plotted for \texttt{pass\_flag=True} (accepted) versus \texttt{pass\_flag=False} (rejected). Vertical dashed lines mark the corresponding decision thresholds used by the filter. Accepted synopses concentrate at higher SBERT similarity, while toxicity and contradiction remain near zero for most samples.}
  \label{fig:ragmovieaug_popularity}
\end{figure}

Figure~\ref{fig:ragmovieaug_popularity} (left panel) shows that the filter is not rejecting at random.
The SBERT similarity distribution for $\mathsf{pass}=1$ is clearly shifted to the right relative to $\mathsf{pass}=0$,
meaning accepted synopses are systematically closer (in embedding space) to the anchor metadata.
Intuitively, a synopsis for an action sci-fi anchor that actually discusses sci-fi action themes and matches the genre or tag cues
tends to land in the $\mathsf{pass}=1$ region, whereas generic storylines or genre-mismatched narratives tend to fall into $\mathsf{pass}=0$.

Table~\ref{tab:critic_score} summarizes critic statistics for accepted versus rejected candidates.
The relevance critics are clearly discriminative: accepted synopses achieve a mean cosine similarity of $0.52$ to the full anchor,
whereas rejected ones average $0.35$; tag-level similarity exhibits a similar separation ($0.51$ vs.\ $0.40$).
By contrast, length distributions are nearly identical (around 64--65 words), suggesting the filter is not simply privileging longer outputs.
Moreover, toxicity and contradiction are typically near zero and rarely trigger rejections under anchor-constrained prompting.
Together, these patterns indicate that the dominant failure mode is \emph{semantic drift} (insufficient anchor alignment) rather than unsafe
or internally inconsistent content.

\begin{table}[t]
  \centering
  \caption{LLMGenQC toy study on 500 movies: critic statistics for LLM-generated synopses, split by pass/fail decision.
  We report the mean and standard deviation (sd) for SBERT similarity to the full anchor ($c^{\text{rel}}$), similarity to
  tag-only anchor ($c^{\text{tag}}$), and length in words ($c^{\text{len}}$).}
  \label{tab:critic_score}
  \begin{tabular}{lccc}
    \toprule
    Group & $c^{\text{rel}}$ (mean$\pm$sd) & $c^{\text{tag}}$ (mean$\pm$sd) & $c^{\text{len}}$ (mean$\pm$sd) \\
    \midrule
    Rejected ($\mathsf{pass}=0$) & $0.35 \pm 0.10$ & $0.40 \pm \,$+0.09 & $63.9 \pm 15.0$ \\
    Accepted ($\mathsf{pass}=1$) & $0.52 \pm 0.09$ & $0.51 \pm \,$0.07 & $65.4 \pm 10.9$ \\
    \bottomrule
  \end{tabular}
\end{table}

\begin{figure}[t]
  \centering
  \includegraphics[width=0.25\linewidth]{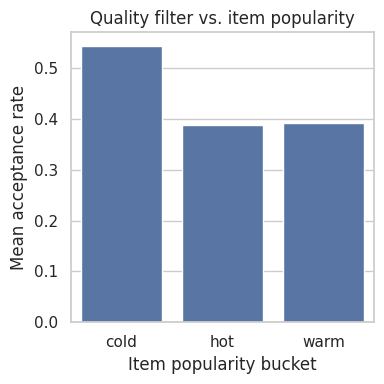}
  \caption{\texttt{LLMGenQC} acceptance rate by item popularity bucket in the 500-movie toy study.
  Cold items (few interactions) retain a larger fraction of synthetic synopses than warm or hot items,
  indicating that the critic-based filter does not disadvantage the long tail.}
  \label{fig:ragmovieaug_popularity_acc}
\end{figure}

\paragraph{Popularity buckets and long-tail behavior.}
We also analyzed how the quality filter behaves across item popularity buckets.
Each movie was assigned to a cold, warm, or hot bucket based on its number of interactions in MovieLens.
For each bucket, we computed the mean fraction of candidates that passed the critic thresholds.
As shown in Figure~\ref{fig:ragmovieaug_popularity_acc}, cold items have the highest acceptance rate (0.54),
whereas warm and hot items exhibit lower acceptance rates around 0.39.
This indicates that the quality filter does not starve the long tail of synthetic descriptions; instead,
it remains slightly more permissive for under-exposed movies, aligning to inject high-quality synthetic text
primarily where human metadata tends to be sparse. Finally, note that this analysis concerns retention of synthetic metadata under QC, rather than downstream exposure in recommendations.

\paragraph{Take-away.}
Overall, this toy study demonstrates that \texttt{LLMGenQC} provides a practical and reproducible LLM-as-judge mechanism
for auditing LLM-generated metadata in recommender benchmarks.
The released critic scores and pass/fail labels enable transparent filtering, sensitivity analysis, and alternative reweighting
strategies such as density-ratio estimation, without re-running the full recommendation experiments.

\subsection{RQ6: Ablation Study}

\begin{figure}[ht!]
     \centering
     \includegraphics[width=1.0\textwidth]{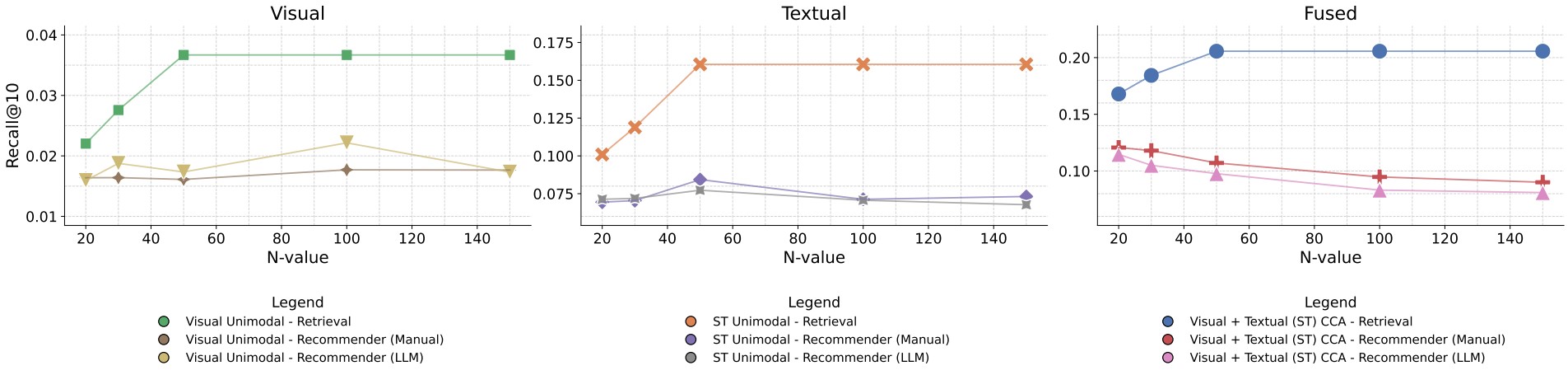}
     \caption{The Recall vs. N chart for different methods and sources.}
     \label{fig_dotchart}
 \end{figure}

Figure \ref{fig_dotchart} illustrates how Recall@10 values evolve as the recommendation set size (N-value) increases across three representation settings, including \textbf{Visual}, \textbf{Textual}, and \textbf{Fused} (Visual + Textual CCA).
It should be noted that the SentenceTransformer variant of textual modality and the CCA method for fusion were employed to shape a proof of concept of the discussions.
The three strategy stages, including \textbf{Retrieval}, \textbf{Manual Recommender}, and \textbf{LLM-based Recommender} were the target of the experiment.
Accordingly, across all subplots, the retrieval-stage recall consistently improves with larger N, indicating that expanding the candidate pool increases the likelihood of retrieving relevant items.
While the Visual-only setting shows relatively low recall overall (compare the first and middle figure), with modest gains as N increases, highlighting the limited discriminative power of purely visual features.
The Textual setting achieves substantially higher recall, and both recommender-based approaches outperform pure retrieval, with the LLM recommender generally yielding the strongest trend.
The Fused setting delivers the best performance across the board, demonstrating clear benefits from combining visual and textual information. However, in the re-ranking stages, increasing \(N\) beyond a moderate range can \emph{reduce} Recall@10 for some settings (notably fused CCA), suggesting that larger candidate pools may introduce harder negatives that degrade the final top-\(K\) selection unless the re-ranker is calibrated for longer lists.

\paragraph{User-embedding ablation (Table~\ref{tab_recommendation_stage_avg}).}
Table~\ref{tab_recommendation_stage_avg} evaluates three user-embedding construction methods used in the retrieval stage: (i) a random embedding baseline, (ii) an average embedding over high-rated items, and (iii) a temporally weighted embedding that upweights recent interactions.
The results show that the temporal embedding yields the most reliable top-\(N\) candidate set for downstream LLM recommendation, improving ranking quality compared to a static average profile.
The random baseline performs substantially worse, confirming that meaningful user representations are necessary for effective retrieval.
Overall, this ablation supports using temporal user embeddings as the default in the main experiments, while all variants remain available via the released configuration schema for reproducibility and extension.

\noindent\textbf{Reproducibility of the audit.}
To ensure full transparency, we release all critic scores (SBERT anchor similarity, NLI contradiction,
toxicity, perplexity, length) and the final accept/reject label as columns in the synthetic synopsis file. A companion notebook reproduces Table~\ref{tab:critic_score} and Figure~\ref{fig_dotchart} end-to-end, including thresholding and
acceptance-rate analysis by popularity bucket.

\begin{table*}[!ht]
    \centering
    \caption{Average Recall and Recommendation Stage Results (Accuracy Metrics). We use the same color-based notation and columns as in Table~\ref{tab:retrieval_stage_final}.
    The best results are taken from Tables~\ref{tab:retrieval_stage_final} and \ref{tab:recommendation_stage_final}.}
    \label{tab_recommendation_stage_avg}
    \resizebox{\textwidth}{!}{%
    \begin{tabular}{lcc|cc|cccc|c}
    \toprule
        \textbf{Name} & \multicolumn{2}{c|}{\textbf{Uni/Multimodal}} & \multicolumn{2}{c|}{\textbf{Fusion}} & \multicolumn{4}{c|}{\textbf{Metrics}} & \multirow{2}{*}{\textbf{$\Delta_{ndcg}$}} \\
    \cmidrule(lr){2-3}\cmidrule(lr){4-5}\cmidrule(lr){6-9}
         & Visual & Text & PCA & CCA & Recall (temp) & NDCG (temp) & Recall (avg) & NDCG (avg) & \\
    \midrule
        \multicolumn{9}{l}{\textbf{Retrieval Stage}} \\
    \midrule
        (Best) Visual+Textual (ST) & \checkmark & \checkmark &      & \checkmark & 0.205637 & 0.252080 & 0.180762 & 0.174588 & $+571.6\%$ \\
    \midrule
        Average (ST) & \checkmark & \checkmark & \checkmark & \checkmark & \best{0.126937} & \best{0.143082} & \best{0.10424875} & \best{0.09652225} & $+281.2\%$ \\
        Average (OpenAI) & \checkmark & \checkmark & \checkmark & \checkmark & 0.1012713 & 0.1056025 & 0.0769490 & 0.0695398 & $+181.4\%$ \\
        Average (Llama3.0) & \checkmark & \checkmark & \checkmark & \checkmark & 0.0896470 & 0.1054925 & 0.0713208 & 0.0690150 & $+181.1\%$ \\
        Average (All) & \checkmark & \checkmark & \checkmark & \checkmark & \secondbest{0.1198045} & \secondbest{0.1341640} & \secondbest{0.0957160} & \secondbest{0.0889310} & $+257.4\%$ \\
    \midrule
        \multicolumn{9}{l}{\textbf{Recommendation Stage -- Manual Pipeline}} \\
    \midrule
        (Best) Visual+Textual (ST) & \checkmark & \checkmark &      & \checkmark & 0.107084 & 0.268102 & 0.095418 & 0.174363 & $+495.5\%$ \\
    \midrule
        Average (ST) & \checkmark & \checkmark & \checkmark & \checkmark & \best{0.062722} & \best{0.151360} & \best{0.048413} & \best{0.091438} & $+236.2\%$ \\
        Average (OpenAI) & \checkmark & \checkmark & \checkmark & \checkmark & 0.039368 & 0.101269 & 0.033289 & 0.068101 & $+124.9\%$ \\
        Average (Llama3.0) & \checkmark & \checkmark & \checkmark & \checkmark & 0.035219 & 0.099198 & 0.030246 & 0.065384 & $+120.3\%$ \\
        Average (All) & \checkmark & \checkmark & \checkmark & \checkmark & \secondbest{0.051702} & \secondbest{0.131727} & \secondbest{0.042383} & \secondbest{0.083869} & $+192.6\%$ \\
    \midrule
        \multicolumn{9}{l}{\textbf{Recommendation Stage -- LLM-based Pipeline}} \\
    \midrule
        (Best) Visual+Textual (ST) & \checkmark & \checkmark & & \checkmark & 0.097654 & 0.246567 & 0.089355 & 0.159758 & $+427.1\%$ \\
    \midrule
        Average (ST) & \checkmark & \checkmark & \checkmark & \checkmark & \best{0.056856} & \best{0.145850} & \best{0.044908} & \best{0.085151} & $+211.8\%$ \\
        Average (OpenAI) & \checkmark & \checkmark & \checkmark & \checkmark & 0.046062 & 0.103145 & 0.032384 & 0.066209 & $+120.5\%$ \\
        Average (Llama3.0) & \checkmark & \checkmark & \checkmark & \checkmark & 0.034305 & 0.097747 & 0.026053 & 0.062965 & $+109.0\%$ \\
        Average (All) & \checkmark & \checkmark & \checkmark & \checkmark & \secondbest{0.051421} & \secondbest{0.129342} & \secondbest{0.039644} & \secondbest{0.080262} & $+176.5\%$ \\
    \bottomrule
    \end{tabular}
    }
\end{table*}

\section{Conclusions}
We introduced a multi-modal recommendation resource that fuses textual signals from \acp{LLM} with trailer-derived visual embeddings under a unified \ac{RAG} pipeline. Our data augmentation step addresses sparse metadata, while \ac{CCA}-based fusion often outperforms simple concatenation in both recall and coverage. Further, re-ranking via an LLM can boost nDCG by exploiting richer context. These findings underscore the benefits of combining textual and visual features for cold-start and novelty-focused recommendations. Looking ahead, we hope this resource will encourage more sophisticated multimodal approaches, including audio embeddings or advanced cross-modal alignment, to tackle domain-specific challenges where item metadata is limited. \vspace{1mm}

\noindent\textbf{Limitations and Future Work.} While our work introduces an open multimodal benchmark and toolkit for Retrieval-Augmented Generation (RAG) in recommendation, several limitations remain:

\begin{itemize}[leftmargin=1.5em,topsep=0.3em]
    \item \textbf{Single-Domain Focus:} Our experiments are restricted to the MovieLens dataset, limiting generalizability to other domains (e.g., music, e-commerce, fashion) where multimodal signals and metadata sparsity may present different challenges. Evaluating the framework on additional, diverse datasets would strengthen claims about robustness and utility.

    \item \textbf{Audio and Rich Multimodality Underexplored:} Although the dataset supports audio features, our current experiments focus only on textual and visual modalities. Integrating audio descriptors or experimenting with other modalities (e.g., subtitles, user reviews) could further illuminate the value and challenges of true multimodal RAG.

    \item \textbf{Limited Baseline Comparisons:} We primarily benchmark against unimodal baselines and our own fusion variants. Comprehensive evaluation against state-of-the-art multimodal recommenders, including recent LLM/RAG-based models and more sophisticated fusion or alignment strategies (e.g., cross-attention, gated fusion), would provide a clearer picture of strengths and weaknesses. We do not implement all contemporary multimodal and LLM/RAG baselines in this paper. Many such baselines introduce additional data requirements and substantial computational footprints, which were beyond the scope and resources of the current study. Accordingly, the experimental focus is on controlled comparisons between unimodal pipelines and modular fusion variants under multiple backbone choices. Implementing additional RAG-based recommenders on top of RAG-VisualRec is a promising direction for future work, and the released code is intended to support exactly these extensions.

    \item Finally, we observe that LLM-based re-ranking can improve nDCG by leveraging richer contextual signals, but it is not a universal remedy and its effect depends on the candidate pool quality. We therefore interpret our cold-start-related findings through \emph{item-side long-tail exposure proxies} (e.g., Coverage and LongTailFrac), and leave dedicated user-side cold-start protocols for future work.

    \item \textbf{Design Principle Positioning:} Our primary contribution is infrastructural—a reproducible multimodal RAG resource, rather than algorithmic novelty. Nonetheless, future versions could incorporate more advanced fusion, retrieval, and generation strategies to push the boundaries of multimodal recommendation research.

\end{itemize}

\noindent
In summary, while \RAGBench~addresses the acute need for transparent, auditable multimodal RAG resources, future work should expand to additional modalities, datasets, and more competitive baselines, and integrate formal robustness and adversarial benchmarks. We invite the community to build on and extend this resource for broader, more robust, and fairer multimodal recommender system research.

Last but not least, while our \texttt{LLMGenQC} toy study provides an automatic, critic-based audit of LLM-generated synopses, we do not conduct a dedicated human evaluation of text quality. Combining our released critic
scores with targeted human judgments are an interesting direction for future work and would further validate the suitability of LLM-generated metadata for sensitive application domains~\cite{deldjoo2025toward}.  Another promising direction is to investigate this line of work in an agentic setup, where agents perform multi-step retrieval and tool use~\cite{maragheh2025future}.
We also plan to explore the feasibility of visual RAG in security-related applications (e.g., performing retrieval over image metadata) and to systematically assess its robustness against multimodal adversarial threats, including stealthy visual attacks and poisoning strategies that can manipulate retrieval and downstream generation outcomes~\cite{nazary2025stealthy,nazary2025poison}.

\bibliographystyle{abbrv}
\bibliography{refs}

\end{document}